\documentclass[twocolumn,twocolappendix]{aastex631}

\usepackage{physics}
\usepackage{amsmath}
\usepackage{tabu}
\usepackage{booktabs}
\usepackage{subfigure}
\usepackage{hyperref}
\usepackage{ulem}
\usepackage{enumerate}
\usepackage{CJK}

\def\sout{\leavevmode \bgroup \ULdepth=-.55ex \ULset}

\newcommand{\sect}{Section~}

\newcommand{\figu}{Figure~}
\newcommand{\figus}{Figures~}
\newcommand{\tab}{Table~}

\newcommand{\eq}{Equation~}

\newcommand{\athena}{\texttt{Athena++}}
\newcommand{\kokkos}{\texttt{Kokkos}}

\shorttitle{Accretion of BH in Ellipticals}
\shortauthors{Guo et al.}
\graphicspath{{./}{figures/}}

\begin{document}
\begin{CJK*}{UTF8}{gbsn}

\title{Toward Horizon-scale Accretion onto Supermassive Black Holes in Elliptical Galaxies}

\correspondingauthor{Minghao Guo}
\email{mhguo@princeton.edu}

\author[0000-0002-3680-5420]{Minghao Guo (郭明浩)}
\affiliation{Department of Astrophysical Sciences, Princeton University, Princeton, NJ 08544, USA}

\author[0000-0001-5603-1832]{James M. Stone}
\affiliation{School of Natural Sciences, Institute for Advanced Study, 1 Einstein Drive, Princeton, NJ 08540, USA}
\affiliation{Department of Astrophysical Sciences, Princeton University, Princeton, NJ 08544, USA}

\author[0000-0003-2896-3725]{Chang-Goo Kim}
\affiliation{Department of Astrophysical Sciences, Princeton University, Princeton, NJ 08544, USA}

\author[0000-0001-9185-5044]{Eliot Quataert}
\affiliation{Department of Astrophysical Sciences, Princeton University, Princeton, NJ 08544, USA}

\begin{abstract}
We present high-resolution, three-dimensional hydrodynamic simulations of the fueling of supermassive black holes in elliptical galaxies from a turbulent medium on galactic scales, taking M87* as a typical case. The simulations use a new GPU-accelerated version of the \athena\ AMR code, and span more than 6 orders of magnitude in radius, reaching scales similar to the black hole horizon. The key physical ingredients are radiative cooling and a phenomenological heating model. 
We find that the accretion flow takes the form of multiphase gas at radii less than about a kpc. The cold gas accretion includes two dynamically distinct stages: the typical disk stage in which the cold gas resides in a rotationally supported disk and relatively rare chaotic stages ($\lesssim 10\%$ of the time) in which the cold gas inflows via chaotic streams. 
Though cold gas accretion dominates the time-averaged accretion rate at intermediate radii, accretion at the smallest radii is dominated by hot virialized gas at most times. The accretion rate scales with radius as $\dot{M}\propto r^{1/2}$ when hot gas dominates and we obtain $\dot{M}\simeq10^\mathrm{-4}-10^\mathrm{-3}\,M_\odot\,\mathrm{yr^{-1}}$ near the event horizon, similar to what is inferred from EHT observations. The orientation of the cold gas disk can differ significantly on different spatial scales. 
We propose a subgrid model for accretion in lower-resolution simulations in which the hot gas accretion rate is suppressed relative to the Bondi rate by $\sim (r_\mathrm{g}/r_{\rm Bondi})^{1/2}$. Our results can also provide more realistic initial conditions for simulations of black hole accretion at the event horizon scale.
\end{abstract}


\section{Introduction} \label{sec:intro}

Supermassive black holes (SMBHs) are ubiquitous in the nuclei of elliptical galaxies~\citep{Kormendy&Ho2013ARA&A..51..511K}. The fueling of these black holes involves a sizeable dynamic range from megaparsec to milliparsec scales and a variety of physical processes including radiative cooling, heating, turbulence, and magnetic fields.

It remains uncertain how SMBHs in galactic nuclei are fed in detail. For spherically symmetric adiabatic accretion of hot gas ~\citep{Bondi1952MNRAS.112..195B}, the accretion begins from a characteristic (Bondi) radius, ($r_\mathrm{B}=2GM_\bullet/c_\mathrm{s,\infty}^2$, where $G$ is the gravitational constant, $M_\bullet$ is the black hole mass, and $c_\mathrm{s,\infty}$ is the sound speed of ambient gas), where the gravitational energy due to the SMBH becomes comparable to the thermal energy of the gas.
For many low-luminosity active galactic nuclei (AGN), it is believed that the quasi-spherical Bondi flow at large radii transitions at smaller radii to a hot, tenuous radiatively inefficient accretion flow (RIAF), with a radiative cooling time-scale longer than the accretion time-scale ~\citep{Ichimaru1977ApJ...214..840I,Narayan&Yi1995ApJ...452..710N}. 
However, if cooling plays an important role, the cold gas will contribute to the accretion and change the accretion flow substantially by forming a geometrically thin disk or chaotic cold accretion flow~\citep{Li&Bryan2012ApJ...747...26L, Sharma2012MNRAS.420.3174S, Gaspari2013MNRAS.432.3401G}, in which the accretion rate is boosted by several orders of magnitude due to cooling of the hot gas. 

Contemporary event horizon-scale modeling of accretion flows is limited in part by a dependence on ad hoc initial conditions.
Most previous general relativistic magnetohydrodynamic simulations (GRMHD)~\citep{Narayan2012MNRAS.426.3241N,Porth2019ApJS..243...26P,White2020ApJ...891...63W} set the initial condition as a torus in hydrodynamic equilibrium with a specified angular momentum profile ~\citep{Fishbone&Moncrief1976ApJ...207..962F, Kozlowski1978A&A....63..209K}. However, the flow structure can strongly depend on the initial conditions, forming structures such as a magnetically arrested disk (MAD)~\citep{Narayan2003PASJ...55L..69N}. More realistic initial and boundary conditions on galactic scales may help to construct a more consistent model of black hole accretion at smaller radii. Connecting the large and small scales in this way is also critical for developing more physical models of black hole growth and feedback in cosmological simulations that lack the physics or resolution to follow the gas to small radii. 

Among massive galaxies, the giant elliptical galaxy M87 is of particular interest. The Event Horizon Telescope~\citep{M87EHT_I_2019ApJ...875L...1E} presented the first image of event horizon scale plasma in the galaxy M87, deriving a BH mass of $M_\bullet=(6.5\pm0.7)\times10^9\,M_\odot$ (consistent with prior stellar dynamics results; \citealt{Gebhardt2009ApJ...700.1690G,Gebhardt2011ApJ...729..119G}). The corresponding gravitational radius is $r_\mathrm{g}=GM_\bullet/c^2=3.1\times10^{-4}\,\mathrm{pc}$. The mass accretion rate near the event horizon is estimated to be $(3-20)\times 10^{-4}\,M_\odot\,\mathrm{yr^{-1}}$~\citep{M87EHT_VIII_2021ApJ...910L..13E}. 
For M87, the temperature of the ambient hot plasma far from the black hole is $\approx2\times10^7\,\mathrm{K}$~\citep{Russell2015MNRAS.451..588R}, which gives a Bondi radius of $r_\mathrm{B}\approx0.12\,\mathrm{kpc}\sim4\times10^5r_\mathrm{g}$ and a Bondi accretion rate of $\dot{M}\approx0.05\,M_\odot\,\mathrm{yr^{-1}}$, much larger than the accretion rate measured on horizon scales. 
In principle, cooling of hot gas on kpc scales and larger could generate an even larger inflow rate of $\sim10\,M_\odot\,\mathrm{yr^{-1}}$, although observations imply that heating (rather than gas inflow) largely offsets this radiative cooling \citep{Fabian2012}. In addition to the hot virialized plasma, HST observations show that there is a warm ionized $T\sim 10^4\,\mathrm{K}$ gas disk within $\sim40\,\mathrm{pc}$ (e.g., \citealt{Walsh2013ApJ...770...86W}).
Despite these constraints on the gas properties $\sim 100$ pc from the SMBH, the realistic structure and properties of the accretion flow in M87 and related systems are still uncertain. To self-consistently link the accretion rate at galactic scales to the event horizon, a dynamic range of $\sim 6$ orders of magnitude must be covered.

This work is a first step towards building a multi-scale model of the accretion flow onto SMBHs in elliptical galaxies from galactic down to event horizon scales, involving different physics at different scales. 
Extending previous works~\citep{Li&Bryan2012ApJ...747...26L, Gaspari2013MNRAS.432.3401G}, our simulations span approximately six orders of magnitude in radius, reaching all the way down to radii $\ll$ pc. As much as possible, we set our initial conditions at large radii using observations~\citep{Gebhardt2009ApJ...700.1690G,Urban2011MNRAS.414.2101U,Russell2015MNRAS.451..588R}.
The simulations connect the black hole accretion rate at small radii to the physical conditions in the galaxy at larger radii. More broadly, the problems that we address in this work include: the contribution of cold and hot gas to the accretion rate, the radial dependence of cold gas and hot gas, as well as the morphology, angular momentum, and dynamics of the accretion flow.
The flow structure found here at small radii can also be used as more realistic initial conditions for simulations of the black hole accretion disk at event horizon scales, which will help to interpret Event Horizon Telescope observations of M87.

The rest of this article is organized as follows. In \sect\ref{sec:method} we describe the physical model and numerical methods we adopt. \sect\ref{sec:results} presents the results of our simulations. In \sect\ref{sec:disscussion}, we discuss the implications of our results. We conclude in \sect\ref{sec:summary}.

\section{Model and Numerical Methods} \label{sec:method}
We perform hydrodynamic simulations using a performance portable version of the \athena~\citep{Stone2020ApJS..249....4S} code implemented using the \kokkos\ library \citep{2021CSE....23e..10T}. \athena\ provides a variety of reconstruction methods, Riemann solvers, and integrators for solving the equations of hydrodynamics. 
The static mesh refinement (SMR) in \athena\ enables us to achieve a high resolution and good performance over an extremely large dynamic range.
We perform the simulations on Cartesian grids to avoid the severe time-step restriction and discontinuities at the pole inherent in 3D spherical-polar coordinates. In addition, there is no a priori symmetry axis in our problem, making Cartesian grids particularly suitable. 

In our simulations, we adopt the piecewise linear (PLM) reconstruction method, the Harten-Lax-van Leer-Contact (HLLC) Riemann solver, and the RK2 time integrator to solve the hydrodynamic equations. 
We also tested the piecewise parabolic (PPM) reconstruction method and the RK3 time integrator, and the results are very similar, although the higher-order methods tend to be less stable, leading to problematic cells with unphysically large temperatures or velocities. Thus our production runs adopt second-order methods.
The source terms (gravity, cooling, and heating) are included by operator splitting.

In this study, we solve
\begin{eqnarray}
\frac{\partial \rho}{\partial t}+\nabla \cdot(\rho \boldsymbol{v}) &=&0, \\
\frac{\partial \rho \boldsymbol{v}}{\partial t}+\nabla \cdot\left(P\mathbf{I}+\rho \boldsymbol{v} \boldsymbol{v}\right) &=&\rho\boldsymbol{g}, \\
\frac{\partial E}{\partial t}+\nabla \cdot\left[\left(E+P\right) \boldsymbol{v}\right] &=&\rho\boldsymbol{g}\cdot\boldsymbol{v}-Q_{-}+Q_{+},\label{eq:energy_eq}
\end{eqnarray}
where $\rho$ is the gas density, $\boldsymbol{v}$ is the velocity, $P$ is the gas pressure,
\begin{equation}
    P=\frac{\rho k_\mathrm{B} T}{\mu m_\mathrm{H}},
\end{equation}
with mean molecular weight $\mu=0.618$ by assuming the gas is fully ionized,
$\boldsymbol{g}$ is the gravitational acceleration, $E$ is the total energy density,
\begin{equation}
    E=E_\mathrm{int}+E_\mathrm{kin}=\frac{P}{\gamma-1}+\frac{1}{2}\rho v^2,
\end{equation}
with the gas adiabatic index $\gamma=5/3$, $Q_{-}$ is the volume cooling rate, and $Q_{+}$ is the heating rate. 
The implementations of initial and boundary conditions, gravitational field, radiative cooling, and heating are presented below.

\begin{figure}[ht!]
    \centering
    \includegraphics[width=\linewidth]{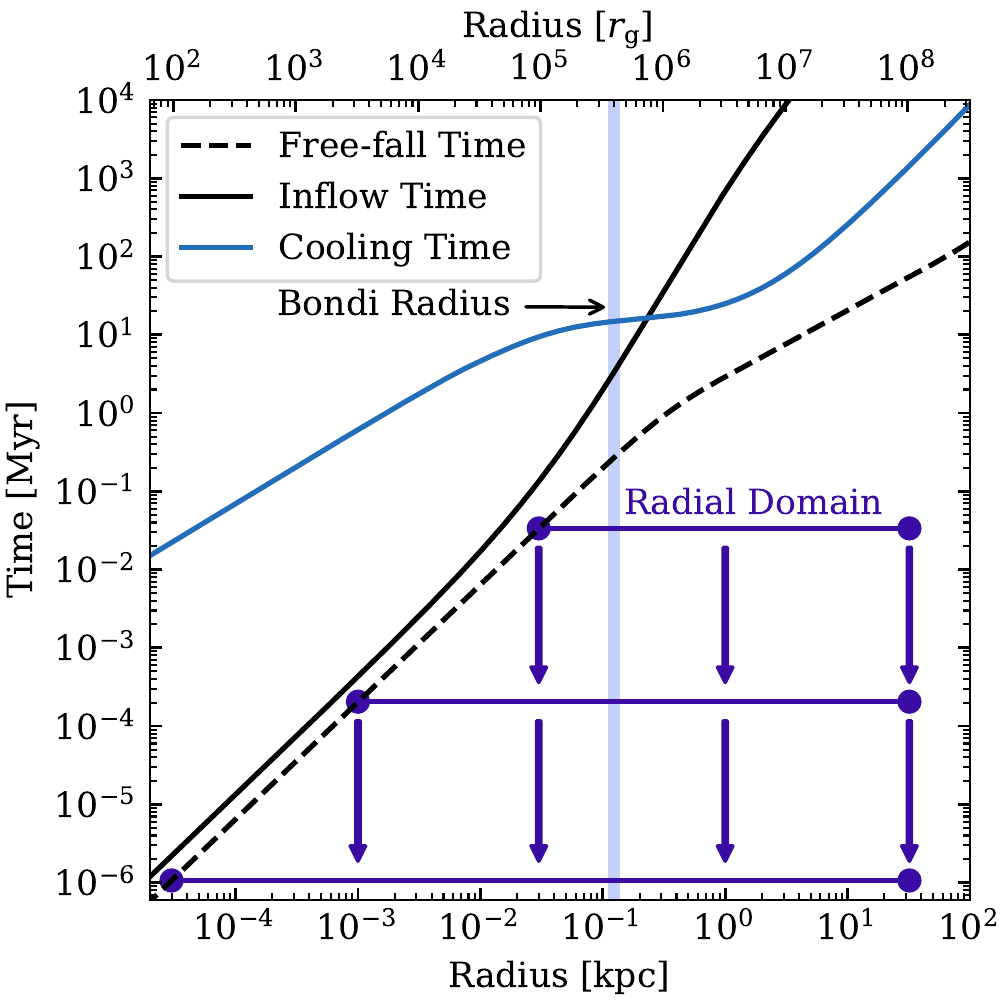}
    \caption{Free-fall time, gas inflow time (assuming a spherical Bondi accretion flow), and gas cooling time as functions of radius. On kpc scales, the gas cooling time is larger than the free-fall time by several times but smaller than the gas inflow time, enabling thermal instabilities to fully develop. The shaded region marks the Bondi radius $r_\mathrm{B}$. The violet lines illustrate the radial domain of the simulations and the arrows indicate that we restart the larger-scale simulation with a smaller inner radius. \label{fig:init_profile}}
\end{figure}

\subsection{Initial and Boundary Conditions}
We investigate the accretion physics in an elliptical galaxy embedded in hot plasma.
For the initial conditions of gas, we assume a flat cored entropy profile, following \citet{Martizzi2019MNRAS.483.2465M}, 
\begin{equation}
    K=\frac{K_0}{2}(1+x^\xi),
\end{equation}
where the normalized radius is defined as $x=r/r_0$.
The gas pressure is 
\begin{equation}
    P=Kn^\gamma,
\end{equation}
where $n=\rho/(\mu m_H)$ is the number density.
Then the density profile is obtained by solving hydrostatic equilibrium,
\begin{equation}
    \frac{K_0}{2}\frac{\dd}{\dd x}\left[(1+x^\xi)n(x)^\gamma\right]+\rho\frac{\dd\Phi}{\dd x}=0,
\end{equation}
where $\Phi$ is the gravitational potential.
In this work, we set $r_0=2\,\mathrm{kpc}$, $K_0=7\,\mathrm{keV\,cm^2}$, and $\xi=1.1$ to match the X-ray observations of the Virgo Cluster in the work of \citet{Urban2011MNRAS.414.2101U}.
The normalization of the density profile is set as $n(x=1)=0.1\,\mathrm{cm^{-3}}$ by taking the observed density fitting from the observations of M87 by~\citet{Russell2015MNRAS.451..588R}. By adopting these parameters, the radial profiles of density and temperature are similar to observations of the galaxy M87 as well as fairly general to be applicable to similar systems.

Some important time-scales of the initial condition are shown in \figu\ref{fig:init_profile}. The free-fall time ranges from roughly 10 Myr at 10 kpc to 1 yr at 0.03 pc ($100 r_\mathrm{g}$). On kpc scale, the gas cooling time $t_\mathrm{cool}$ ($\sim 10\,\mathrm{Myr}$) is larger than the free-fall time $t_\mathrm{ff}$ but by no more than a factor of 10, i.e., $t_\mathrm{cool}/t_\mathrm{ff}\lesssim 10$. Furthermore, assuming a Bondi accretion rate, the cooling time is smaller than the gas inflow time $t_\mathrm{inflow}$. This implies that cooling is dynamically important; indeed, as we shall see, thermal instability develops in this region.

The simulations adopt a cubic box of size $64\,\mathrm{kpc}$ in each direction and cover a radial domain from $32$ kpc down to $30$ mpc or smaller, using static mesh refinement to span 6 orders of magnitude in scale. The center of the potential is placed at the center of the simulation box.
The root grid is a cube of $128^3$ cells, which is divided into $4\times4\times4$ mesh blocks, with each mesh block being a cube of $32^3$ cells. 
At each level of mesh refinement, we double the resolution by refining the central $2\times2\times2$ parent mesh blocks into $4\times4\times4$ child mesh blocks.
This enables us to retain a resolution of $1/64 < \Delta x/x < 1/32$ over the whole simulation domain. 
In the simulations, we typically employ 10 to 20 levels of static mesh refinement. In one case for the largest zoom-in simulation, we use 25 levels with inner radius $r_\mathrm{in}=1\,\mathrm{mpc}(\approx3r_\mathrm{g})$ and the finest resolution of $\sim 0.05 r_\mathrm{g}$.
In one high-resolution simulation for testing convergence, we double the resolution to $1/128 < \Delta x/x < 1/64$ everywhere.

For the outer boundary conditions, we fix the outer region of the box ($r>r_\mathrm{out}=32\,\mathrm{kpc}$) to be the initial hydrostatic equilibrium solution. This allows us to enforce spherical symmetry at the outer boundaries. Since the outer radius $r_\mathrm{out}$ is sufficiently large, the fixed outer boundaries have little influence on the flow at small scales.
For the inner boundaries, we apply a vacuum sink region with $r<r_\mathrm{in}$ by evacuating the gas within the region and resetting a negligible fixed density $n_\mathrm{sink}=5\times10^{-3}\,\mathrm{cm^{-3}}$, temperature $T_\mathrm{sink}=2\times10^5\,\mathrm{K}$, and zero velocity every time step. This sink region avoids artificial overpressure bounces and correctly reproduces the Bondi accretion solution with robust numerical stability.

To deal with the excessively severe time-step restriction at small scales, we first evolve the simulations with a relatively larger inner radius $r_\mathrm{in,old}$. Then we restart the simulation with a smaller inner radius $r_\mathrm{in,new}$. The new cells in the simulation domain are initialized with the values used for the boundary conditions.
The evolution time for each run is $\gtrsim10^4$ free-fall time at the inner boundary. In \figu\ref{fig:init_profile}, we illustrate the radial domain of the simulations with different $r_\mathrm{in}$.
When restarting with a smaller inner radius, to avoid the sudden change of inner boundary conditions and the unphysical shocks therefrom, we gradually shrink the radius of the inner boundary by
\begin{equation}
    r_\mathrm{in}(t)=r_\mathrm{in,new}\frac{t}{t_\mathrm{trans}}+r_\mathrm{in,old}\left(1-\frac{t}{t_\mathrm{trans}}\right),
\end{equation}
where the transition time-scale, $t_\mathrm{trans}$, is a few times longer than the local dynamical time at $r_\mathrm{in}$ so that the change is nearly adiabatic. We tested the effects of the transition time on the accretion flow. The results show little difference for different $t_\mathrm{trans}$ as long as $t_\mathrm{trans}$ is long enough to avoid the strong unphysical shocks.

The simulations are initialized with random perturbations to model turbulent motions, which is crucial in seeding thermal instability and shaping the accretion flow. 
We seed thermal instabilities using two types of perturbations, i.e., density perturbations (associated with lower angular momentum) and velocity perturbations (higher angular momentum).
In the first type, we set an isobaric density perturbation, in which the velocity of the perturbations is relatively small. This initial perturbation means a relatively smaller dispersion of angular momentum of the gas.
In the second type, we set a divergence-free Kolmogorov-like velocity perturbation. 
The amplitude of velocity is subsonic, with $\sigma_v\sim 100\,\mathrm{km\,s^{-1}}$, similar to the turbulent velocities in elliptical galaxies~\citep{dePlaa2012A&A...539A..34D,Sanders&Fabian2013MNRAS.429.2727S}.
This perturbation induces a larger root-mean-square angular momentum, but still with a very small net angular momentum. The kinetic energy in the initial velocity perturbation is roughly $1\%$ of the internal energy.
We do not continuously drive the turbulence during the computation but only seed the velocity perturbation in the initial condition. In practice, we set the velocity perturbation only on large scales, with wavelength $3.2\,\mathrm{kpc}<\lambda<8\,\mathrm{kpc}$, and let it naturally cascade to smaller scales during the evolution. Note that the nonlinear cascade time $\sim \lambda/\sigma_v \sim 10^8$ yrs which is comparable to the duration of our simulation.

\subsection{Gravitational Field}
We adopt a spherically symmetric gravitational field with the gravitational acceleration being
\begin{equation}
    \boldsymbol{g}(\boldsymbol{r})=-\frac{\dd\Phi}{\dd\boldsymbol{r}}=-\frac{GM(r)\boldsymbol{r}}{r^3},
\end{equation}
where $M(r)$, the mass enclosed in radius $r$, is the sum of the mass of the SMBH, stars, and dark matter,
\begin{equation}
    M(r)=M_\bullet+M_\star(r)+M_\mathrm{DM}(r).
\end{equation}
We do not include the self-gravity of the gas.
For M87, the \citet{M87EHT_I_2019ApJ...875L...1E} derived a SMBH mass of $M_\bullet=(6.5\pm0.7)\times10^9\,M_\odot$, corresponding to a gravitational radius of $r_\mathrm{g}=3.1\times10^{-4}\,\mathrm{pc}\approx0.3\,\mathrm{mpc}$, consistent with the dynamical inferred SMBH mass by \citet{Gebhardt2009ApJ...700.1690G}. For the stars and dark matter, \citet{Gebhardt2009ApJ...700.1690G} presented radial profiles using observations of stellar light and kinematics.
We model the mass of stars using a broken power law for density, similar to the NFW profile~\citep{Navarro_NFW_1997ApJ...490..493N},
which gives a profile of mass as
\begin{equation}
    M_\star(r)=M_\mathrm{s}\left[\ln(1+\frac{r}{r_\mathrm{s}})-\frac{r}{r_\mathrm{s}+r}\right],
\end{equation}
where we set $r_\mathrm{s}=2\,\mathrm{kpc}$ and $M_\mathrm{s}=3\times10^{11}\,M_\odot$. For the dark matter, we also adopt the NFW profile with $r_\mathrm{DM}=60\,\mathrm{kpc}$ and $M_\mathrm{DM}=3\times10^{13}\,M_\odot$.
The potential at the inner boundary is softened by
\begin{equation}
    \Phi=-\frac{GM(r)}{\sqrt{r^2+s^2}},
\end{equation}
where we set $s^2=r_\mathrm{a}^2\exp[-(r/r_\mathrm{a})^2]$ with $r_\mathrm{a}=0.5r_\mathrm{in}$.

\subsection{Cooling and Heating}
\label{sec:coolheat}
The cooling function in \eq\ref{eq:energy_eq} takes the form of 
\begin{equation}
    Q_{-}(n,T)=n^2\Lambda(T),
    \label{eq:cooling_func}
\end{equation}
where $\Lambda(T)$ is the emissivity.
For temperatures above $10^{8.15}\,\mathrm{K}$ we use a fit introduced by~\citet{Schneider2018ApJ...860..135S},
\begin{equation}
    \Lambda(T)=10^{-26.065}T^{0.45}\,\mathrm{erg\,s^{-1}\,cm^3}.
\end{equation}
For $T<10^{8.15}\,\mathrm{K}$, we use a linear interpolation of the tabulated cooling rate taken from Table 2 of~\citet{Schure2009A&A...508..751S} for solar metallicity. We include this cooling source term by operator splitting forward Euler method.

Despite the presence of radiative cooling, X-ray observations of cluster cores do not show a sharp drop in temperature or the presence of X-ray lines from cooler gas ~\citep{Vikhlinin2006ApJ...640..691V}, implying that the system is in global thermal equilibrium due to the presence of heating.
The source of heating, unlike cooling, is more uncertain and difficult to model analytically. 
The major source of heating is probably AGN feedback, but with a possible contribution from galaxy mergers, stellar feedback, conductive heating, plasma viscous heating, etc. 
Similar to \citet{Sharma2012MNRAS.420.3174S} and \citet{Gaspari2013MNRAS.432.3401G}, we assume an idealized phenomenological heating prescription, in which the global cooling is balanced by global heating with a heating rate approximately equal to the angle-averaged cooling rate at the same radius,
\begin{equation}
    Q_{+}(\boldsymbol{r},t)=\frac{f_\mathrm{Q}}{4\pi}\int_{0}^{2\pi}\int_{0}^{\pi}Q_{-}[n(\boldsymbol{r},t),T(\boldsymbol{r},t)]\sin\theta\dd\theta\dd\phi,
    \label{eq:heating_func}
\end{equation}
where the normalization factor $f_\mathrm{Q}=0.98$ in practice, slightly smaller than unity.
This heating prescription provides a way to keep a global equilibrium but not a local equilibrium. Locally, due to thermal instability, some dense gas cools quickly and forms cold clumps and filaments, leading to a two-phase structure of the gas. 

We tested the effects of different normalizations of the heating rate. Overall, we find that a higher heating rate tends to suppress the formation of cold gas and therefore decrease the accretion rate. Increasing the heating rate causes the accretion rate to gradually decrease from the pure cooling limit to the adiabatic limit. In fact, if we choose normalization such that the total heating rate equals the cooling rate ($f_Q=1$), there is little cold gas generated, especially during the later stage of the simulation. The heating prescription in Equation \ref{eq:heating_func} occasionally produces artificial overheating when there is considerable cold gas, leading to the formation of unphysically high-temperature gas. So in practice, we do not include the cooling rate from the cold gas ($T<2\times10^6\,\mathrm{K}$) in evaluating the heating in \eq\ref{eq:heating_func}.
We also artificially turn off the heating function for gas with $T>10^9\,\mathrm{K}$.
In the central region, heating plays a less significant role. It is unlikely that heating from AGN feedback operates the same on small scales as it does on larger scales in ellipticals and clusters. For example, jets appear to deposit most of their energy on scales of kpc or greater. For this reason, we elected to turn off heating within 1 pc of the BH, although we find that it does not make a big difference to the results.
In practice, we begin heating after 5 Myr and smoothly increase it from zero to the value given in \eq\ref{eq:heating_func} at 10 Myr in order to accelerate the formation of cold gas~\citep[similar to][]{Gaspari2013MNRAS.432.3401G}. 

\subsection{Floors and Flux Corrections}
Using \athena\,, we evolve the conservative variables (mass, momentum, and total energy) and treat the source terms by operator splitting. 
In practice, however, due to the combination of strong shocks, strong rarefaction, and strong cooling over the large dynamic range, the primitive variables of $\rho$ and $P$ can occasionally reach unphysical values, e.g., they can be negative.
In addition, accurate integration of the cooling term generally requires a time step much smaller than the hydrodynamic time step, imposing significant restrictions on the computational speed. 
Lastly, the cold gas can readily cool to temperatures of $10^4\,\mathrm{K}$, making the scale height of the disk too small to resolve, especially at small radii in the simulations.
Thus instead of imposing a cooling time step limit, we apply floors when calculating the primitive variables from the conserved variables.
For density, we apply both a fixed density floor of $n_\mathrm{floor}=10^{-3}\,\mathrm{cm^{-3}}$ and a radius-dependent density floor of $10^{-3}(r/1\,\mathrm{kpc})^{-1}\,\mathrm{cm^{-3}}$. For temperature, we apply a floor of $T_\mathrm{floor}=2\times10^{5}\,\mathrm{K}$. 
The temperature floors help to resolve the vertical structure of the cold disk so that the dynamical evolution within the disk can be more physically meaningful. In this way, we accelerate the computation significantly and ensure that the model remains stable. We tested the effects of floors by running simulations with a higher $T_\mathrm{floor}$. The density of the cold gas is smaller, but most results, such as accretion rate, are still similar.

Furthermore, we adopt the first-order flux correction algorithm~\citep[described by][]{Lemaster&Stone2009ApJ...691.1092L} to update problematic cells with unphysically large temperatures or velocities.
In this algorithm, we first check the cells after the full update to identify the problematic cells in which the density is smaller than the density floors or the density ratio of adjacent cells is larger than a certain criterion ($\sim10^2$ in practice). Then we roll back, replace the higher-order numerical fluxes at the interface of the problematic cells with spatially first-order fluxes computed using the Local Lax Friedrichs (LLF) solver and update the cells.
This helps to improve the stability of the algorithm and reduces (though does not eliminate) the need for floors. In practice, we find the properties of accretion flow in the normal cells are not affected with or without this method. In general, the correction is required in less than $0.05\%$ of the cells.

The parameters of the simulations and the device type for running them are summarized in \tab\ref{tab:models}.

\begin{deluxetable*}{lcccccrrrrrc}
    \tablenum{1}
    \tablecaption{List of simulations and their parameters. \label{tab:models}}
    \tablewidth{30pt}
    \tablehead{
    \colhead{Model} & \colhead{Cooling} & \colhead{Heating} & \colhead{Turbulence} & \colhead{Root} & \colhead{SMR} & \colhead{Finest} & \colhead{Inner} & \colhead{Evolution} & \colhead{Transition} & \colhead{Restart}  & \colhead{Device}\\[-0.2cm]
    \colhead{Label} & \colhead{Function} & \colhead{Function} & \colhead{Type} & \colhead{Grid} & \colhead{Level} & \colhead{Resolution} &\colhead{Radius\tablenotemark{a}} & \colhead{Time} & \colhead{Time} & \colhead{Time} & \colhead{Type}\\[-0.8cm]
    }
    \decimalcolnumbers
    \startdata
    \textbf{CHD10} & Yes & Yes & Density & $128^3$ & 10 & 0.49 pc & 30 pc & 300 Myr & \multicolumn{1}{c}{--} & \multicolumn{1}{c}{--} & cpu \\
    CHD10-r256 & Yes & Yes & Density & $256^3$ & 10 & 0.24 pc & 30 pc & 50 Myr & \multicolumn{1}{c}{--} & \multicolumn{1}{c}{--} & gpu \\
    CHD15-t030 & Yes & Yes & Density & $128^3$ & 15 & 15 mpc & 1 pc & 2 Myr & 1 Myr & 30 Myr & cpu \\
    CHD15-t090 & Yes & Yes & Density & $128^3$ & 15 & 15 mpc & 1 pc & 2 Myr & 1 Myr & 90 Myr & cpu \\
    \textbf{CHD15-t200} & Yes & Yes & Density & $128^3$ & 15 & 15 mpc & 1 pc & 2 Myr & 1 Myr & 200 Myr & cpu \\
    CHD20-t030 & Yes & Yes & Density & $128^3$ & 20 & 0.48 mpc & 30 mpc & 20 kyr & 10 kyr & 1.5 Myr & cpu \\
    CHD20-t090 & Yes & Yes & Density & $128^3$ & 20 & 0.48 mpc & 30 mpc & 20 kyr & 10 kyr & 2 Myr & cpu \\
    \textbf{CHD20-t200} & Yes & Yes & Density & $128^3$ & 20 & 0.48 mpc & 30 mpc & 20 kyr & 10 kyr & 2 Myr & cpu \\
    CHD25-t200 & Yes & Yes & Density & $128^3$ & 25 & 15 \textmu pc & 1 mpc & 150 yr & 50 yr & 30 kyr & cpu \\
    CHD12 & Yes & Yes & Density & $128^3$ & 12 & 0.12 pc & 1 pc & 100 Myr & \multicolumn{1}{c}{--} & \multicolumn{1}{c}{--} & cpu \\
    CHV10 & Yes & Yes & Velocity & $128^3$ & 10 & 0.49 pc & 30 pc & 300 Myr & \multicolumn{1}{c}{--} & \multicolumn{1}{c}{--} & cpu \\
    CHV15-t100 & Yes & Yes & Velocity & $128^3$ & 15 & 15 mpc & 1 pc & 2 Myr & 1 Myr & 100 Myr & cpu \\
    CHV20-t100 & Yes & Yes & Velocity & $128^3$ & 20 & 0.48 mpc & 30 mpc & 20 kyr & 10 kyr & 2 Myr & cpu \\
    AD10 & No & No & Density & $128^3$ & 10 & 0.49 pc & 30 pc & 200 Myr & \multicolumn{1}{c}{--} & \multicolumn{1}{c}{--} & cpu \\
    AV10 & No & No & Velocity & $128^3$ & 10 & 0.49 pc & 30 pc & 200 Myr & \multicolumn{1}{c}{--} & \multicolumn{1}{c}{--} & cpu \\
    AD15-t150 & No & No & Density & $128^3$ & 15 & 15 mpc & 1 pc & 2 Myr & 1 Myr & 150 Myr & cpu \\
    AV15-t150 & No & No & Velocity & $128^3$ & 15 & 15 mpc & 1 pc & 2 Myr & 1 Myr & 150 Myr & cpu \\
    AD20-t150 & No & No & Density & $128^3$ & 20 & 0.48 mpc & 30 mpc & 10 kyr & 10 kyr & 2 Myr & cpu \\
    AV20-t150 & No & No & Velocity & $128^3$ & 20 & 0.48 mpc & 30 mpc & 10 kyr & 10 kyr & 2 Myr & cpu \\
    A10 & No & No & No & $128^3$ & 10 & 0.49 pc & 30 pc & 300 Myr & \multicolumn{1}{c}{--} & \multicolumn{1}{c}{--} & gpu \\
    C10\tablenotemark{b} & Yes & No & No & $128^3$ & 10 & 0.49 pc & 30 pc & 300 Myr & \multicolumn{1}{c}{--} & \multicolumn{1}{c}{--} & gpu \\
    A13\tablenotemark{c} & No & No & No & $128^3$ & 13 & 0.06 pc & 0.3 pc & 30 Myr & \multicolumn{1}{c}{--} & \multicolumn{1}{c}{--} & cpu \\
    C13\tablenotemark{b}\tablenotemark{c} & Yes & No & No & $128^3$ & 13 & 0.06 pc & 0.3 pc & 30 Myr & \multicolumn{1}{c}{--} & \multicolumn{1}{c}{--} & cpu \\
    CD10\tablenotemark{c} & Yes & No & Density & $128^3$ & 10 & 0.49 pc & 30 pc & 300 Myr & \multicolumn{1}{c}{--} & \multicolumn{1}{c}{--} & cpu \\
    CV10\tablenotemark{c} & Yes & No & Velocity & $128^3$ & 10 & 0.49 pc & 30 pc & 300 Myr & \multicolumn{1}{c}{--} & \multicolumn{1}{c}{--} & cpu \\
    \enddata
    \tablecomments{The models of black hole accretion in this work. The bold simulations are the fiducial models. (1) In the model labels, C stands for cooling, H stands for heating, D stands for density perturbation, V stands for velocity perturbation, and A stands for adiabatic. The digits following the character for the initial conditions indicate the number of refinement levels used in the model. (11) Restart time for the zoom-in simulation from the larger scale simulation. (12) The type of device on which we run the simulation.
    \tablenotetext{a}{The gravitational radius of the SMBH in M87 is $r_\mathrm{g}=0.31\,\mathrm{mpc}$.}
    \tablenotetext{b}{A temperature floor of $T_\mathrm{floor}=10^4\,\mathrm{K}$ is used.}
    \tablenotetext{c}{These models are presented in the Appendix.}}
\end{deluxetable*}

\begin{figure*}[ht!]
    \centering
    \includegraphics[width=\linewidth]{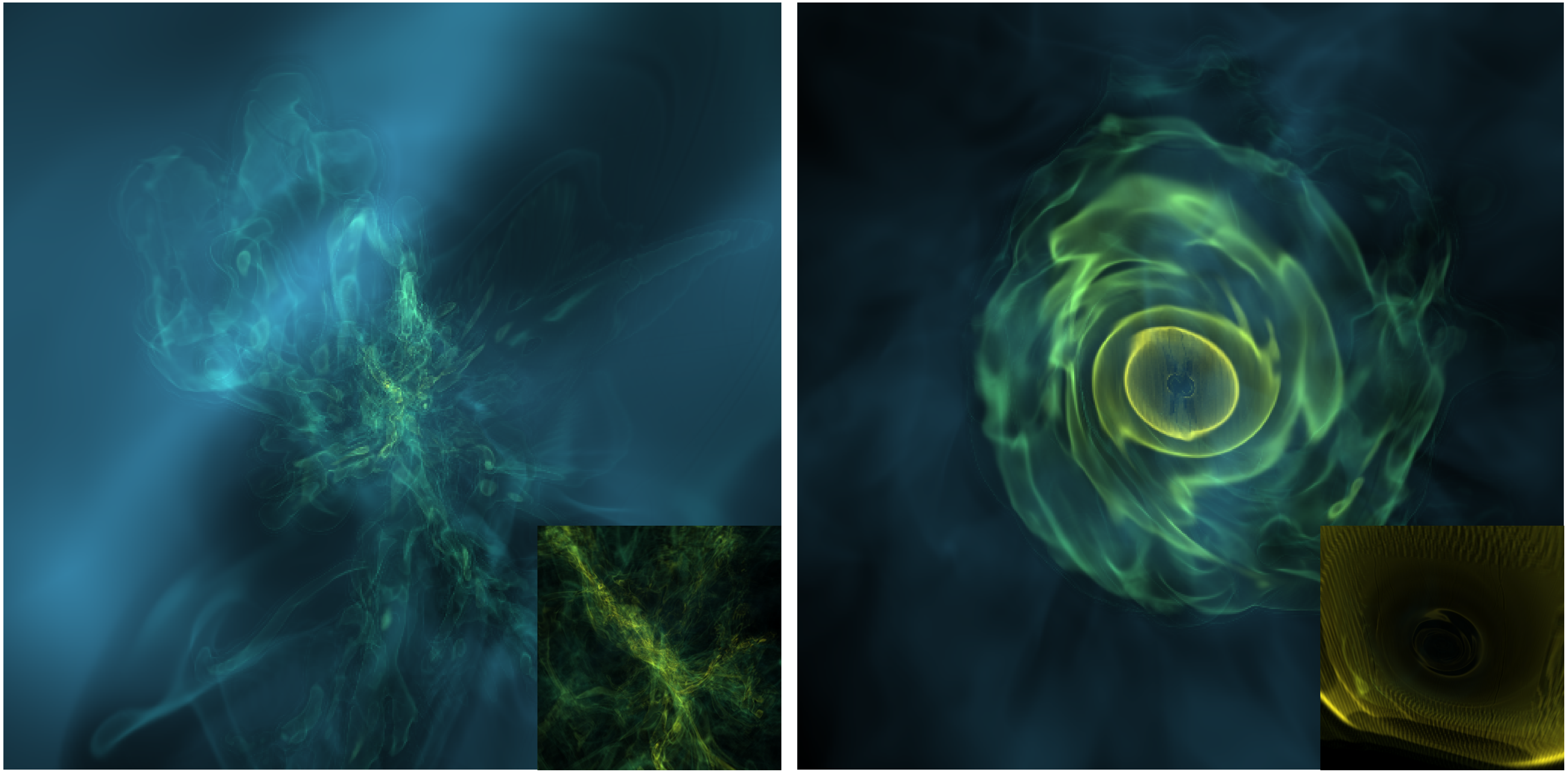}
    \caption{Three-dimensional rendering over a 2$\,\times\,$2 kpc scale region centered on the black hole of chaotic cold accretion (left panel) and cold disk accretion (right panel) emerging in our simulations. This rendering is made using twelve ``layers'' evenly spaced logarithmically in number density between $10^{-3}$ and $10^3\,\mathrm{cm^{-3}}$. A zoom-in of a 200$\,\times\,$200 pc scale region is shown in the bottom right in both images. Usually, a torus structure with size ranging from $\lesssim 100$ pc to nearly 1 kpc is present, depending on the angular momentum. Even with chaotic accretion, there is generally a small torus-like structure within $\sim10\,\mathrm{pc}$. Occasionally, due to collisions of cold clouds and filaments, the torus disappears. \label{fig:fidu_render}}
\end{figure*}

\begin{figure*}[ht!]
    \centering
    \includegraphics[width=\linewidth]{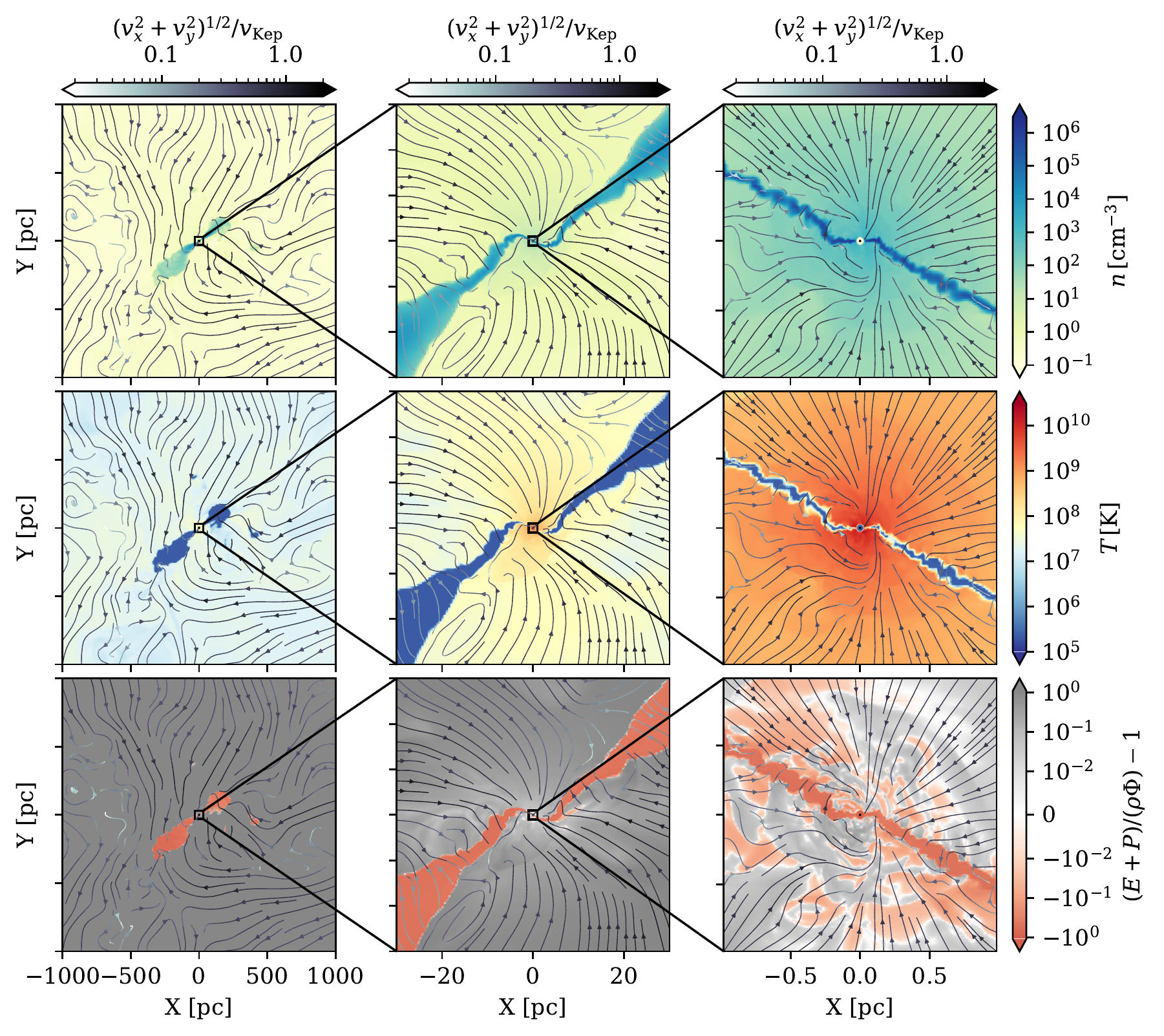}
    \caption{An example slice of density (top), temperature (middle), the Bernoulli parameter $(E+P)/(\rho\Phi)-1$ (bottom), and streamlines of gas velocity through the $z=0$ plane for the disk phase in the simulation with cooling, heating, and density perturbations (Model CHD20-t200, $t=20\,\mathrm{kyr}$). From left to right are zoom-in slices covering $1\,\mathrm{kpc}$, $30\,\mathrm{pc}$, and $1\,\mathrm{pc}$, respectively. There is a cold dense torus of size $r\sim 300\,\mathrm{pc}$ stretching down to the sink cell at very small scale ($\sim 0.03\,\mathrm{pc}$). Note the direction of the disk can differ at different scales. The hot gas is chaotic at the large scale but more laminar at the small scale. \label{fig:fidu_t200_slice}}
\end{figure*}

\begin{figure*}[ht!]
    \centering
    \includegraphics[width=\linewidth]{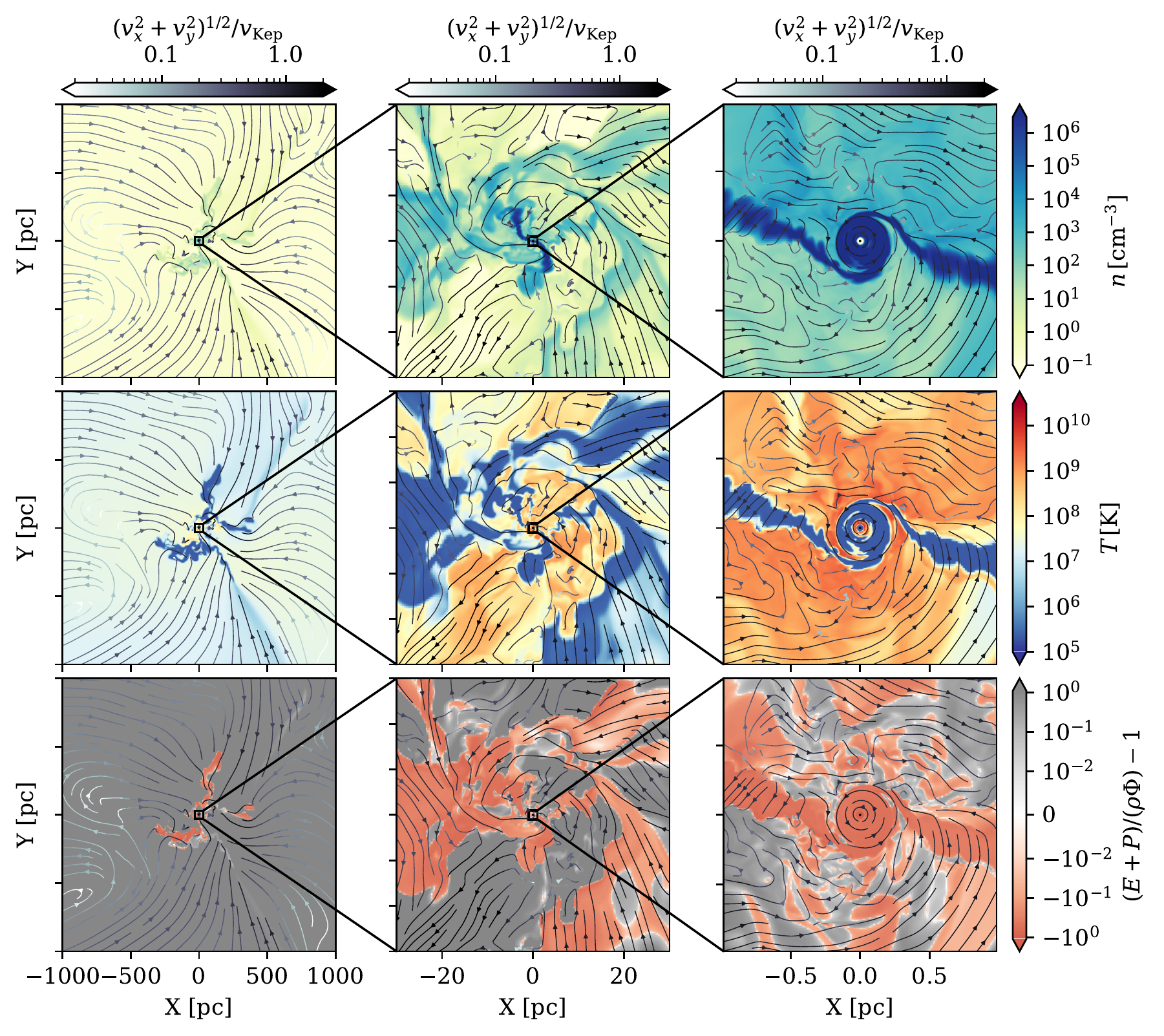}
    \caption{Same as \figu\ref{fig:fidu_t200_slice}, but for the chaotic stage of cold gas accretion (Model CHD20-t030, $t=20\,\mathrm{kyr}$). Though chaotic on large scales, there is still a cold dense torus within $\sim 10\,\mathrm{pc}$ stretching down to the sink cell. The direction of the disk differs at different scales even within $10$ pc. The streamlines show that the hot gas dynamics is much more turbulent on small scales compared to the case with an extended gas disk in \figu\ref{fig:fidu_t200_slice},. \label{fig:fidu_t030_slice}}
\end{figure*}

\begin{figure*}[ht!]
    \centering
    \includegraphics[width=\linewidth]{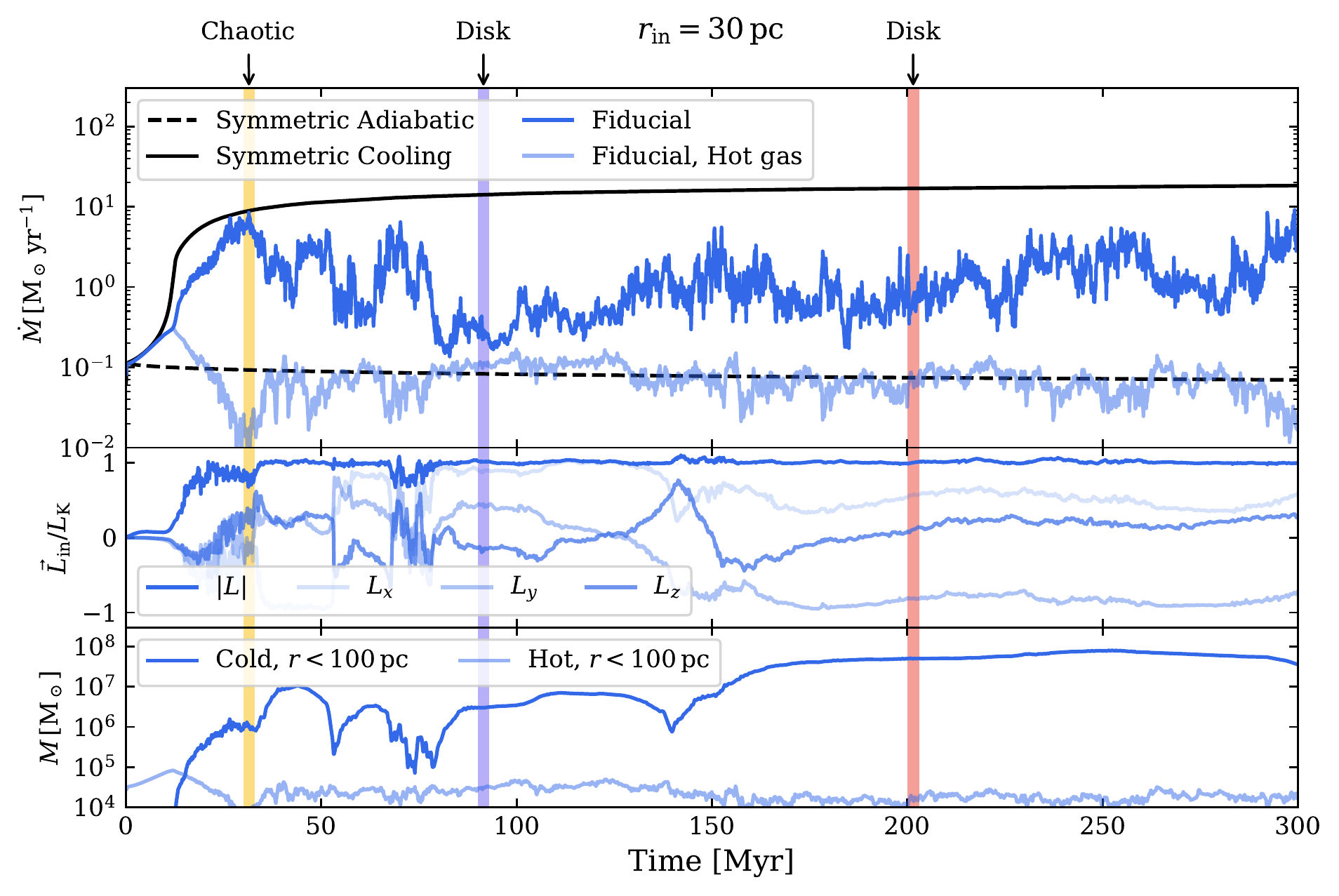}
    \caption{Time evolution of (from top to bottom) mass accretion rate, mass-weighted angle-averaged specific gas angular momentum through the inner boundary, and mass of cold gas for the fiducial simulation (Model CHD10) with cooling, heating, and initial density perturbation. The variables are smoothed with a time bin of 0.1 Myr. The black dashed and solid lines are spherically symmetric simulations. The shaded region marks the time range (30 Myr, 90 Myr, 200 Myr) when we restart the simulation with a smaller $r_\mathrm{in}$. The accretion is turbulent, with $\dot{M}$ roughly fluctuating between the two limits set by the spherically symmetric simulations. The gas is nearly Keplerian near the sink region with the orientation of angular momentum changing frequently.\label{fig:fidu_evo}}
\end{figure*}

\begin{figure*}[ht!]
    \centering
    \includegraphics[width=\linewidth]{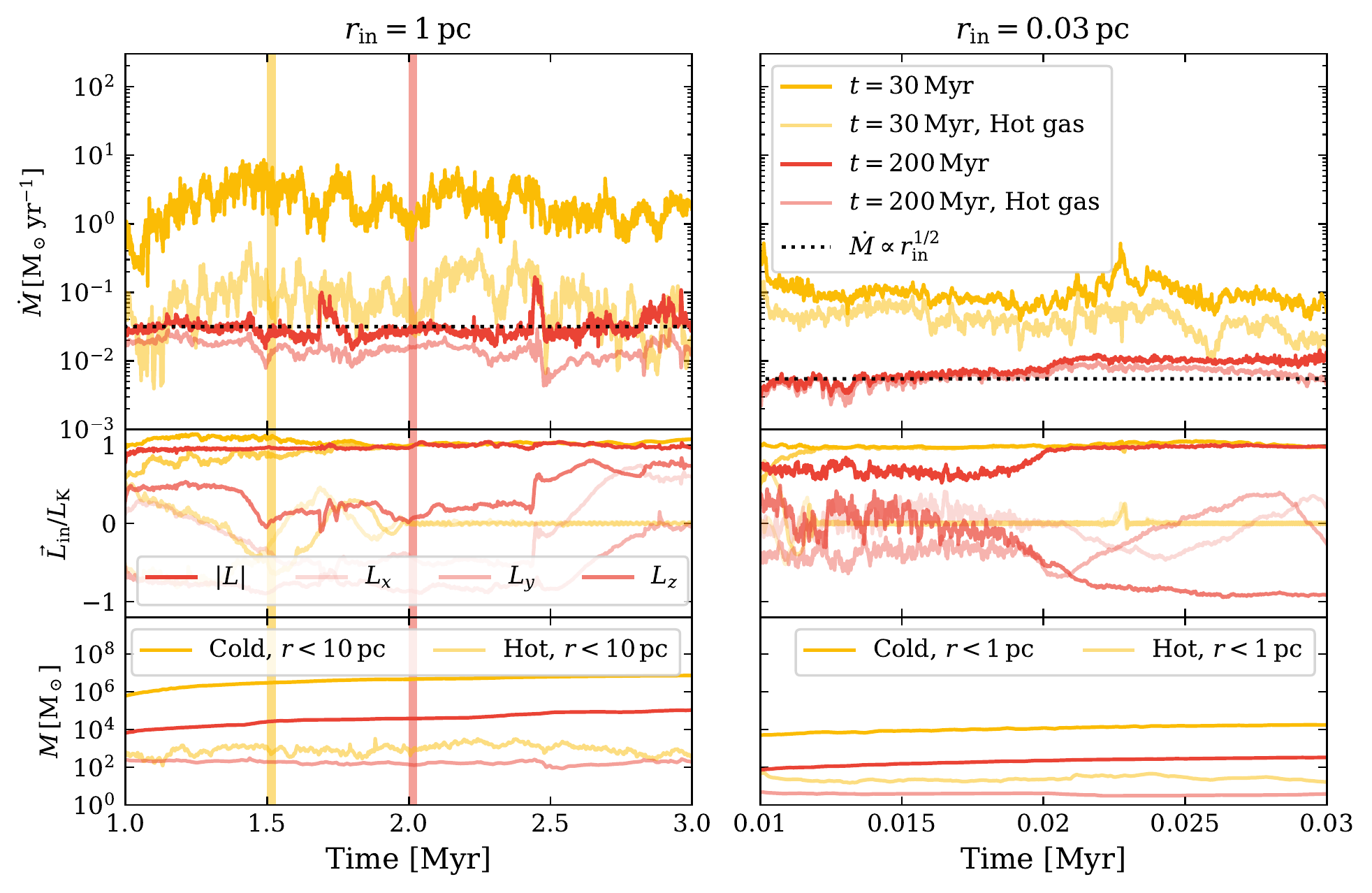}
    \caption{Similar to~\figu\ref{fig:fidu_evo} but for a chaotic case (Models CHD...-t030, yellow) and a disky case (Models CHD...-t200, red) with inner radius $r_\mathrm{in}=1,0.03\,\mathrm{pc}$ The variables are smoothed with a time bin of 1 kyr and 0.01 kyr, respectively. The shaded region marks the time range when we restart the simulation with a smaller $r_\mathrm{in}$. The black dotted lines mark an accretion rate proportional to $r_\mathrm{in}^{1/2}$. The accretion follows $\dot{M}\propto r_\mathrm{in}^{1/2}$. The gas is nearly Keplerian near the sink cell with angular momentum changing frequently. The chaotic accretion and the disky accretion show significant differences. The accretion rate in chaotic accretion tends to be higher than the disky cases. There is more cold gas on small scales for the chaotic cases.\label{fig:fidu_evo_15_20}}
\end{figure*}

\begin{figure*}[ht!]
    \centering
    \includegraphics[width=\linewidth]{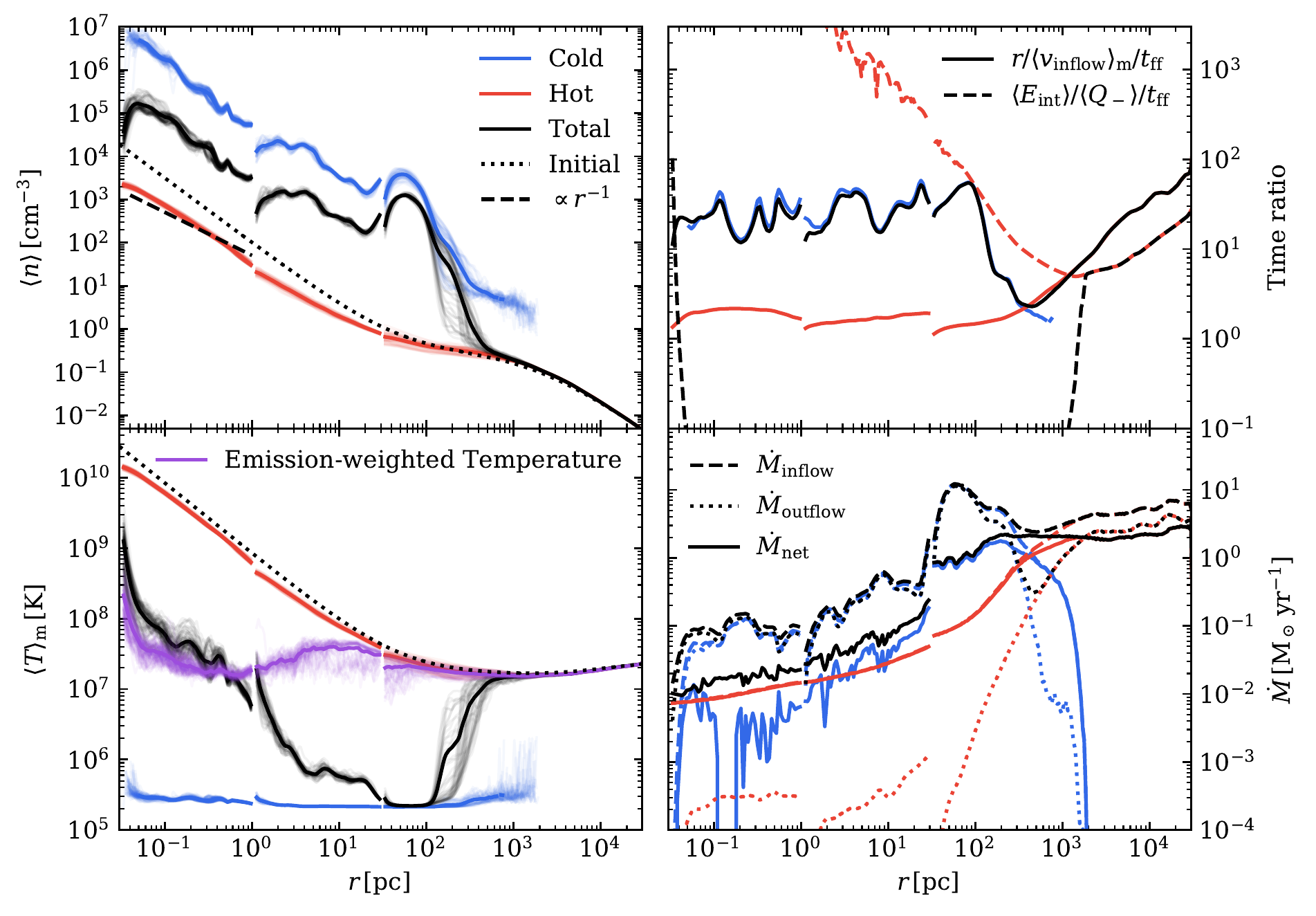}
    \caption{Radial profile of time- and angle-averaged density (top left), temperature (bottom left), inflow and cooling time (top right), and accretion rate (bottom right) for the disk stage of the fiducial suite of simulations CHD10 ($r>30\,\mathrm{pc}$, averaged between 175-225 Myr), CHD15-t200 ($1<r<30\,\mathrm{pc}$, 1.75-2.25 Myr), and CHD20-t200 ($r<1\,\mathrm{pc}$, 20-30 kyr). The lighter lines show the variances of angle-averaged density and temperature within the averaging time range. The emission-weighted temperature is also shown in purple, where we use a range of $0.5-7\,\mathrm{keV}$. For the hot gas, the variables follow power law in radius of $\rho\propto r^{-1}$, $T\propto r^{-1}$, $\dot{M}\propto r^{1/2}$. The cold gas is isothermal with a higher density and large inflow and outflow rate. The emission-weighted temperature profile is flat over a wide range of radii, implying that at small radii, the gas that dominates $\sim$ keV emission is not the hot viral gas but rather the intermediate gas with T $\sim$ keV. \label{fig:fidu_radial}}
\end{figure*}

\begin{figure*}[ht!]
    \centering
    \includegraphics[width=0.9\linewidth]{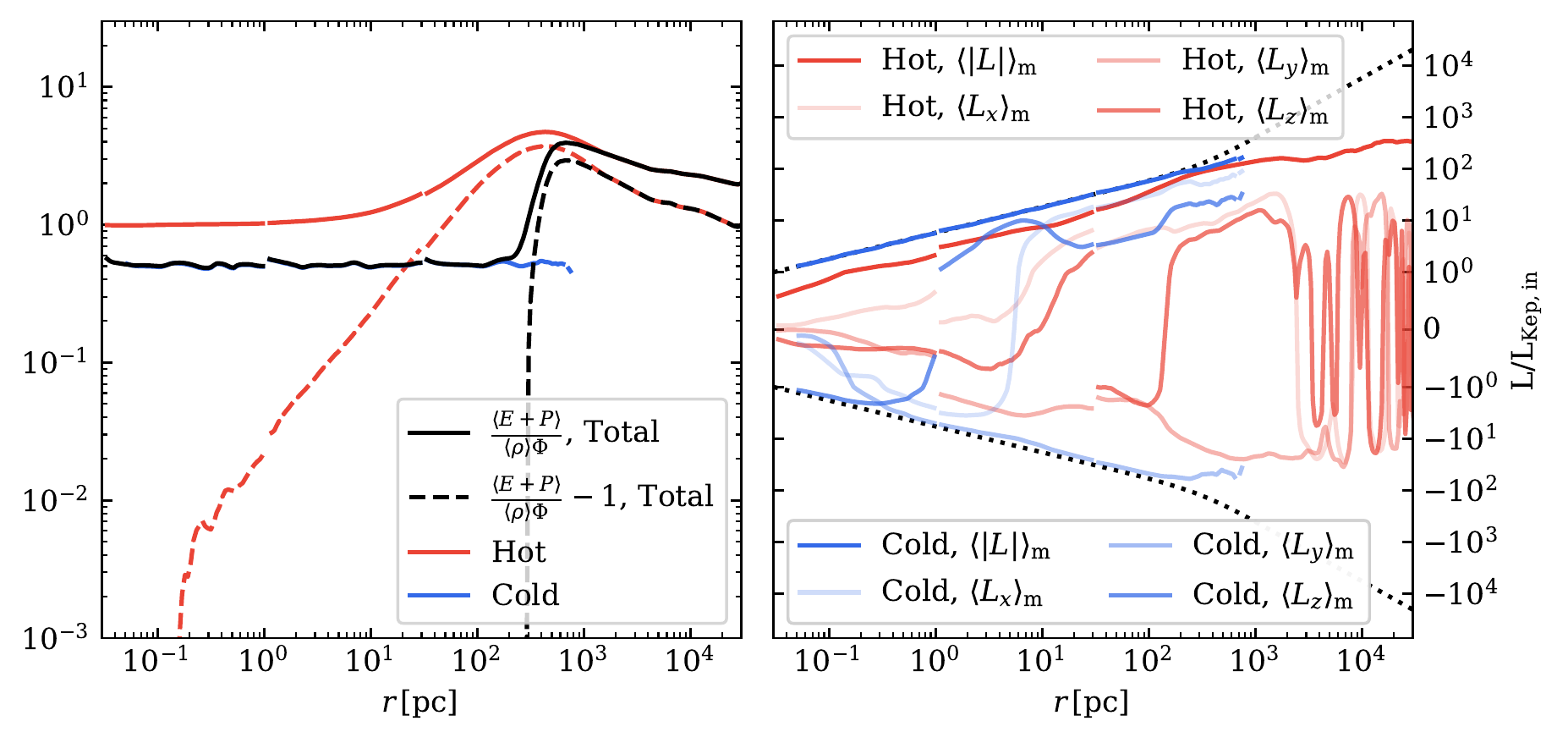}
    \caption{Radial profile of time- and angle-averaged Bernoulli parameter (left) and mass-weighted specific angular momentum (right) for the disk stage of the fiducial suite. The time range for averaging is the same as in \figu\ref{fig:fidu_radial}. The dotted lines are Keplerian angular momentum. The hot gas is slightly unbound and sub-Keplerian. The cold gas is bound in Keplerian motion.\label{fig:fidu_radial_be_am}}
\end{figure*}

\begin{figure*}[ht!]
    \centering
    \includegraphics[width=\linewidth]{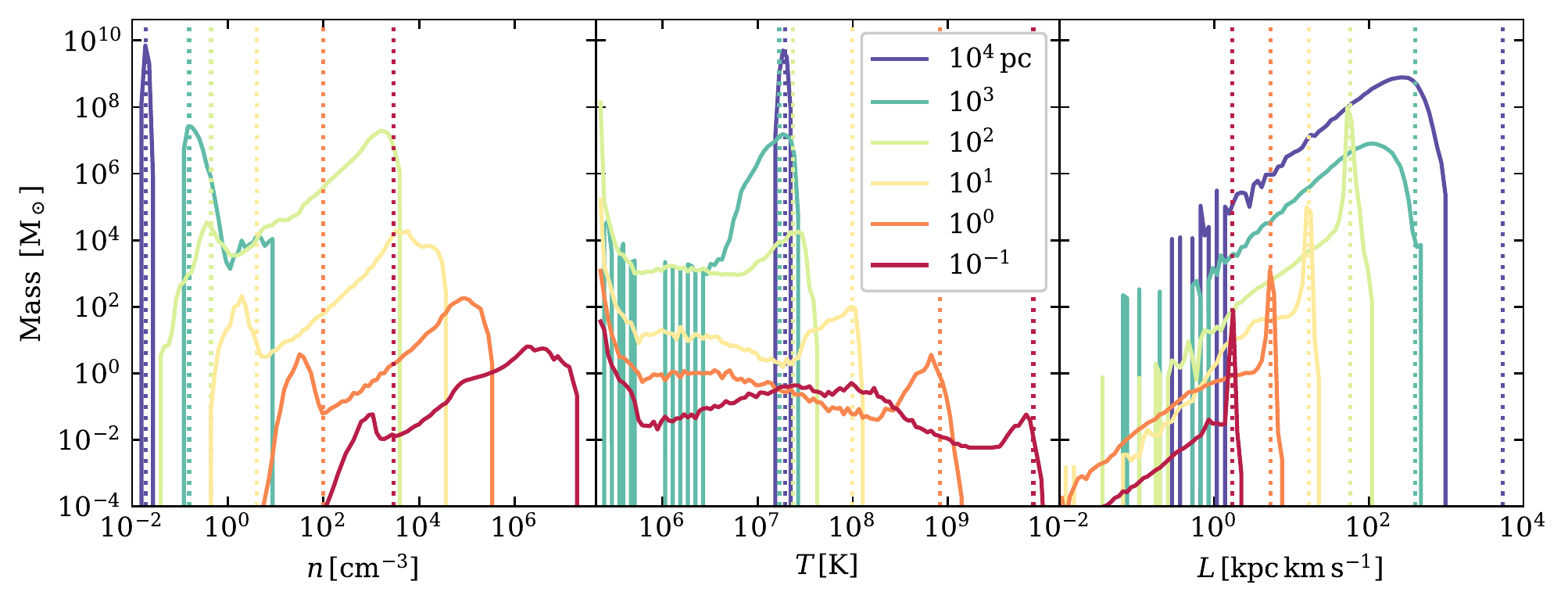}
    \caption{Mass distribution of gas density, temperature, and angular momentum at different radii of the disk stage (Model CHD20-t200). Dashed lines are initial density, initial temperature, and Keplerian angular momentum in the corresponding radius. There is a clearly bimodal distribution of density and temperature within $\sim 1\,\mathrm{kpc}$, indicating that the gas is two-phase. The distribution of specific angular momentum has a long tail. \label{fig:fidu_dist}}
\end{figure*}

\begin{figure}[ht!]
    \centering
    \includegraphics[width=\linewidth]{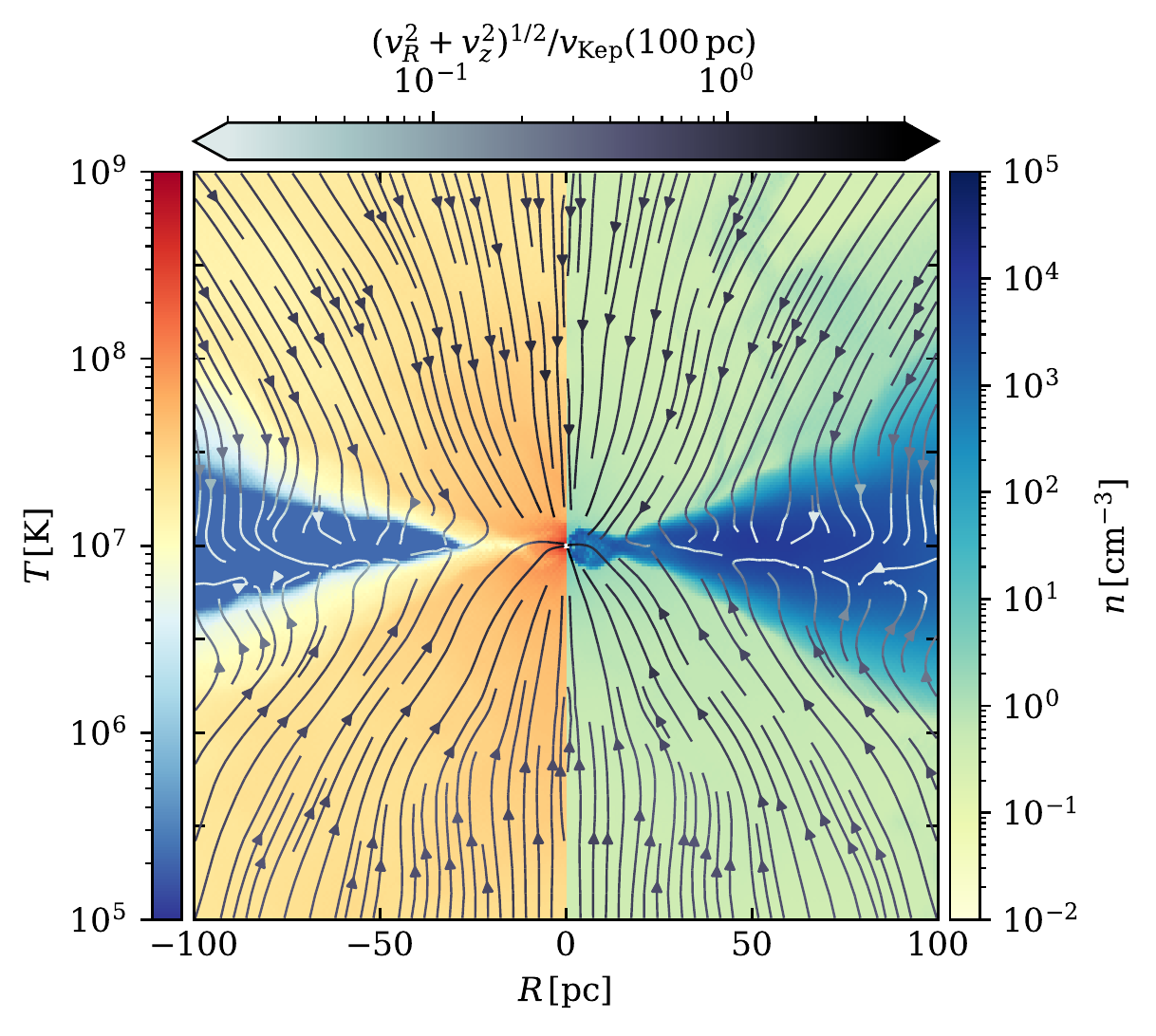}
    \caption{Snapshot of azimuthally-averaged temperature (left) and density (right) for disky stage (Model CHD20-t200) in a coordinate system defined by the angular momentum of the cold gas. The velocity is normalized by the Keplerian velocity at 100 pc. The accretion flow is laminar with smooth streamlines. The accretion is mostly from the polar region and the hot gas both accretes to smaller radii and condenses onto the surface of the cold disk. \label{fig:phi_average}}
\end{figure}

\begin{figure}[ht!]
    \centering
    \includegraphics[width=\linewidth]{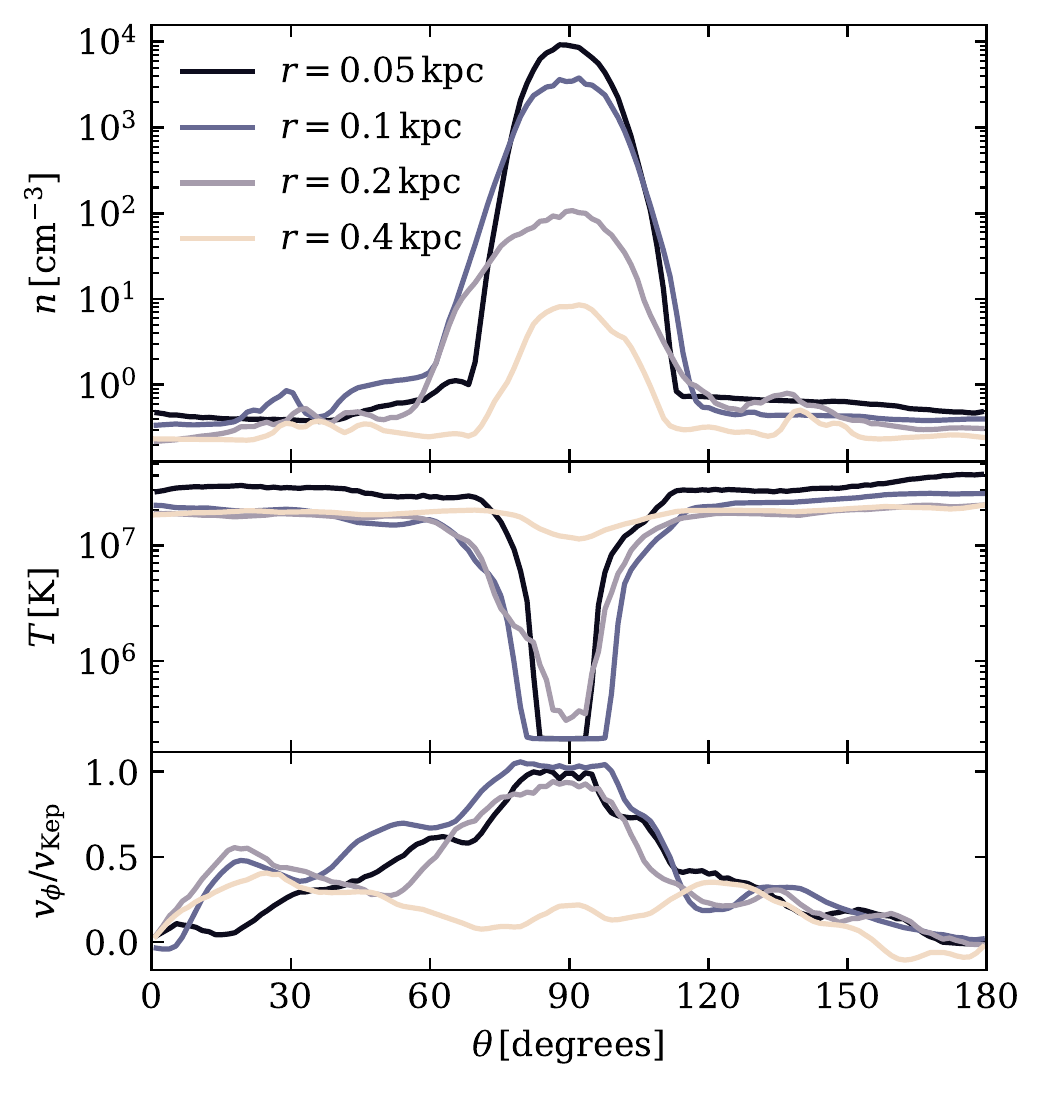}
    \caption{Angular profile of the azimuthally-averaged gas density (top), temperature, (middle), and rotational velocity for the disk stage (Model CHD20-t200) in a coordinate system defined by the angular momentum of the cold gas. There is a clear cold dense Keplerian disk at $\sim 100\,\mathrm{pc}$. \label{fig:angular}}
\end{figure}

\begin{figure*}[ht!]
    \centering
    \includegraphics[width=\linewidth]{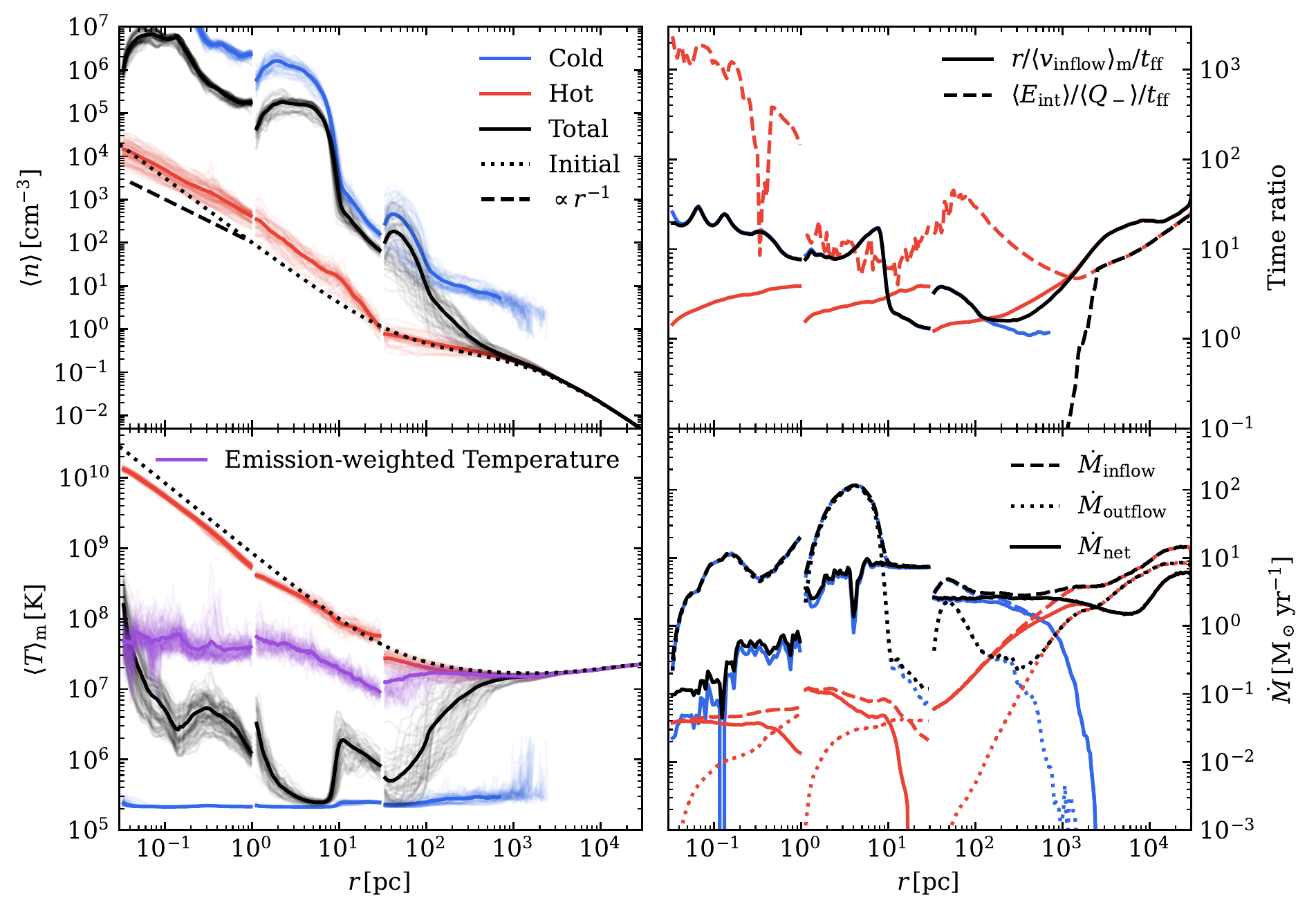}
    \caption{Same as \figu\ref{fig:fidu_radial}, but for the chaotic case. The accretion rate can be as high as $\sim 10\,M_\odot\,\mathrm{yr^{-1}}$ until $\sim 3\,\mathrm{pc}$. For the hot gas, there is a strong outflow that is not present in the disky simulations in \figu\ref{fig:fidu_radial}. \label{fig:fidu_t030_radial}}
\end{figure*}

\begin{figure*}[ht!]
    \centering
    \includegraphics[width=0.9\linewidth]{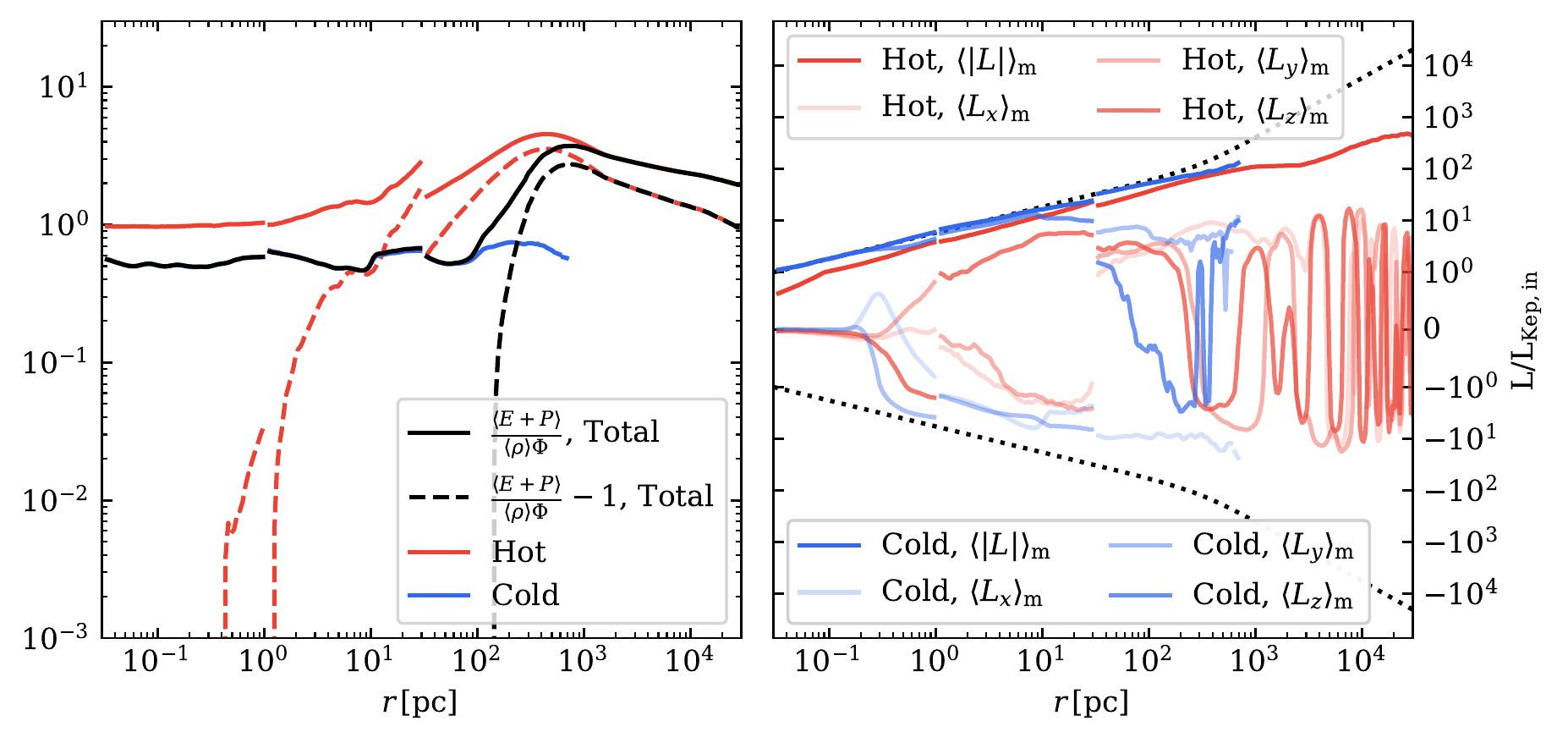}
    \caption{Same as \figu\ref{fig:fidu_radial_be_am}, but for the chaotic case. The hot gas is slightly unbound and sub-Keplerian. The cold gas is bound over the simulation domain and Keplerian within 10 pc. The direction of angular momentum of the cold gas differs on different spatial scales.
    \label{fig:fidu_t030_radial_be_am}}
\end{figure*}

\begin{figure*}[ht!]
    \centering
    \includegraphics[width=\linewidth]{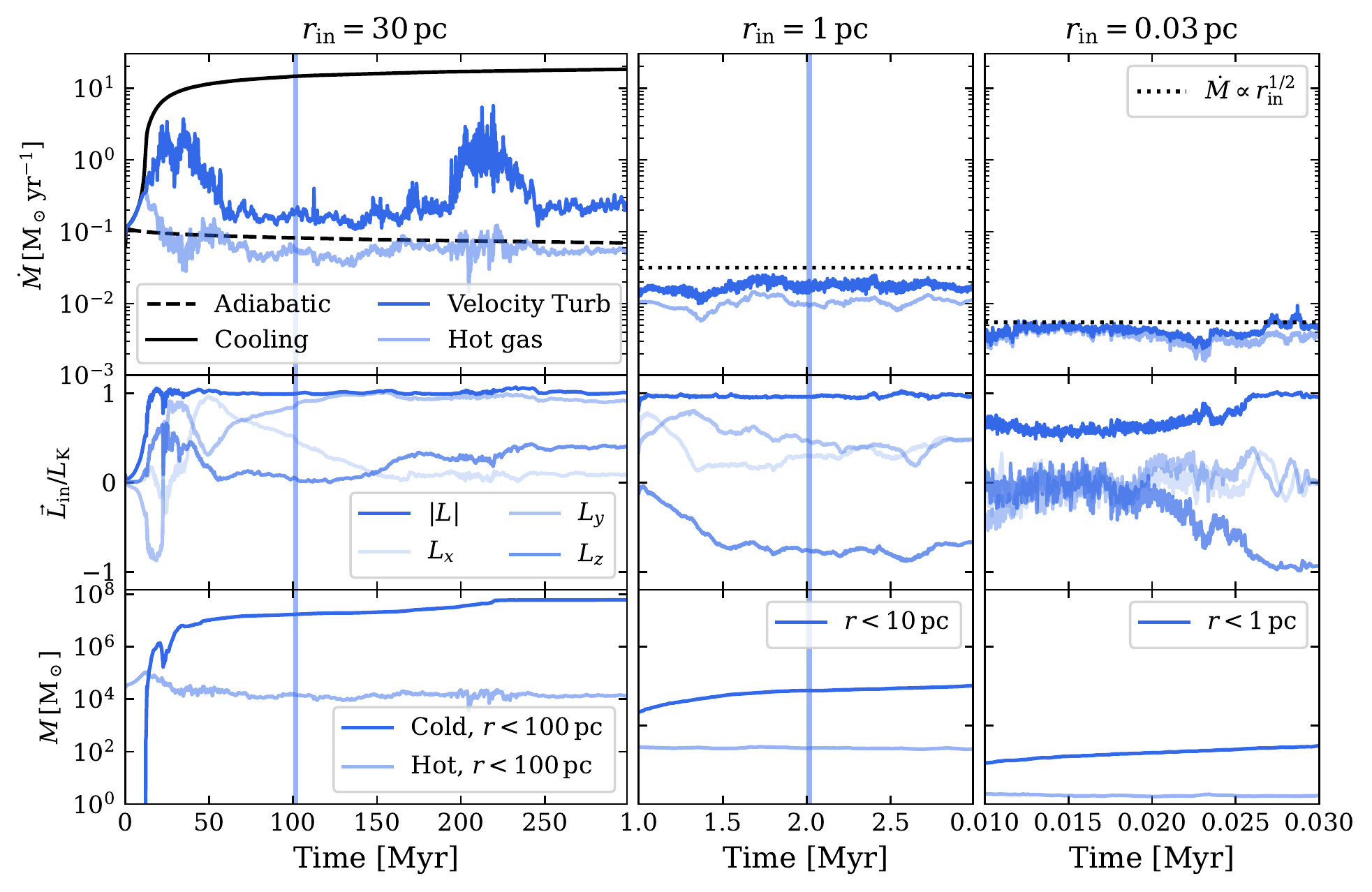}
    \caption{Similar to \figus\ref{fig:fidu_evo} and \ref{fig:fidu_evo_15_20}, but with initial velocity perturbations (Models CHV...), which induces a larger angular momentum. There is a large torus sustained for a long time. The angular momentum changes less frequently than in the fiducial simulations shown in \figus\ref{fig:fidu_evo} and \ref{fig:fidu_evo_15_20}. The accretion rate nonetheless follows a similar scaling to the fiducial suite of simulations. \label{fig:vtur_evo}}
\end{figure*}

\begin{figure*}[ht!]
    \centering
    \includegraphics[width=\linewidth]{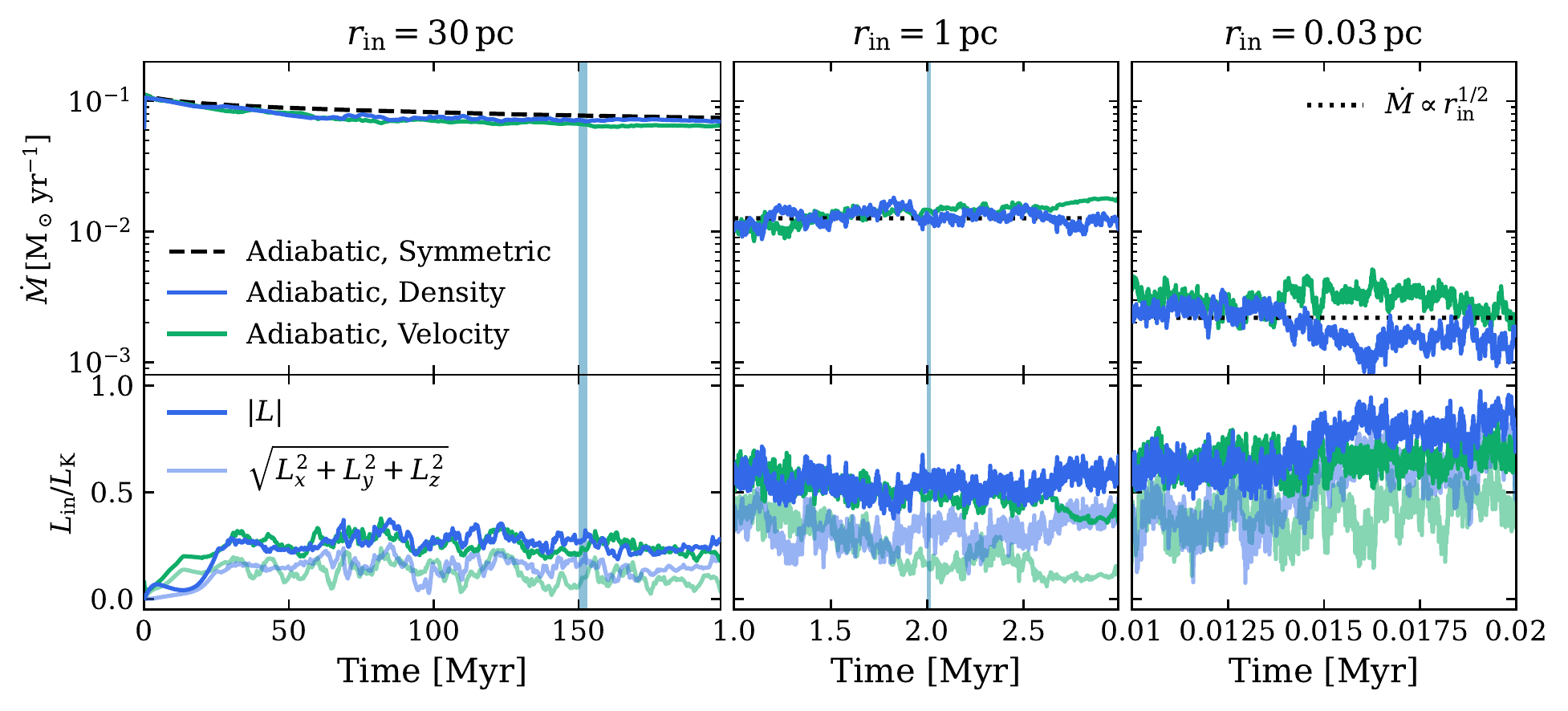}
    \caption{Similar to \figus\ref{fig:fidu_evo} and \ref{fig:fidu_evo_15_20}, but for the adiabatic turbulent simulations (Models AD... and AV...). The black dashed line is the spherically symmetric adiabatic simulations. The accretion follows a clear scaling of $\dot{M}\propto r_\mathrm{in}^{1/2}$. The angular momentum approaches Keplerian, but it is still sub-Keplerian with a relatively large dispersion. There is little difference between simulations with density and velocity perturbations.\label{fig:adb_evo}}
\end{figure*}

\begin{figure*}[ht!]
    \centering
    \includegraphics[width=0.9\linewidth]{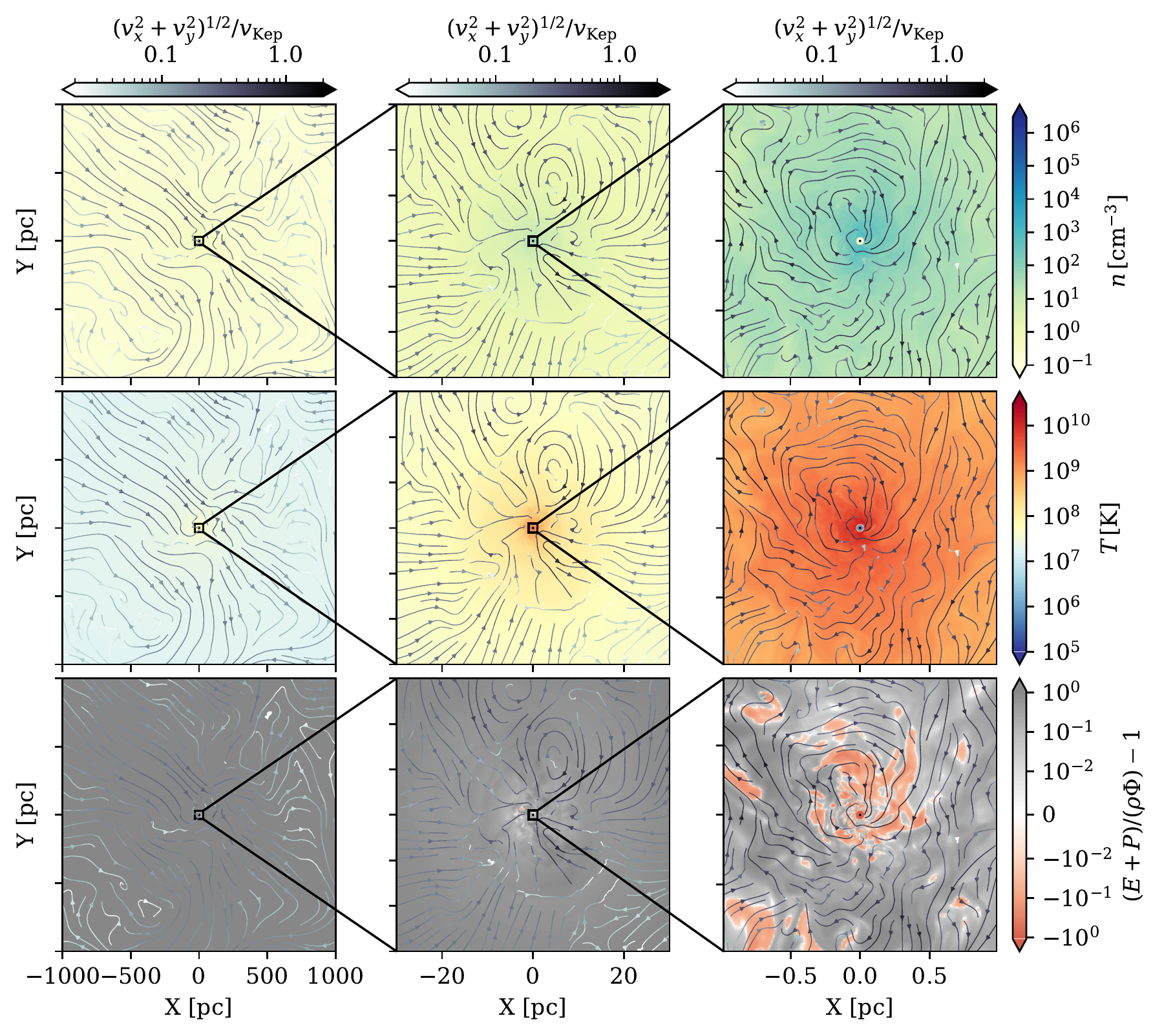}
    \caption{Same as \figu\ref{fig:fidu_t200_slice}, but for the adiabatic simulation with initial density perturbations. The density and temperature show small fluctuations, but the streamlines are very chaotic over the simulation domain. This is very different from the fiducial simulations with radiative cooling, in which the streamlines are laminar on small scales in the presence of a cold accretion disk. \label{fig:adb_slice}}
\end{figure*}

\begin{figure}[ht!]
    \centering
    \includegraphics[width=\linewidth]{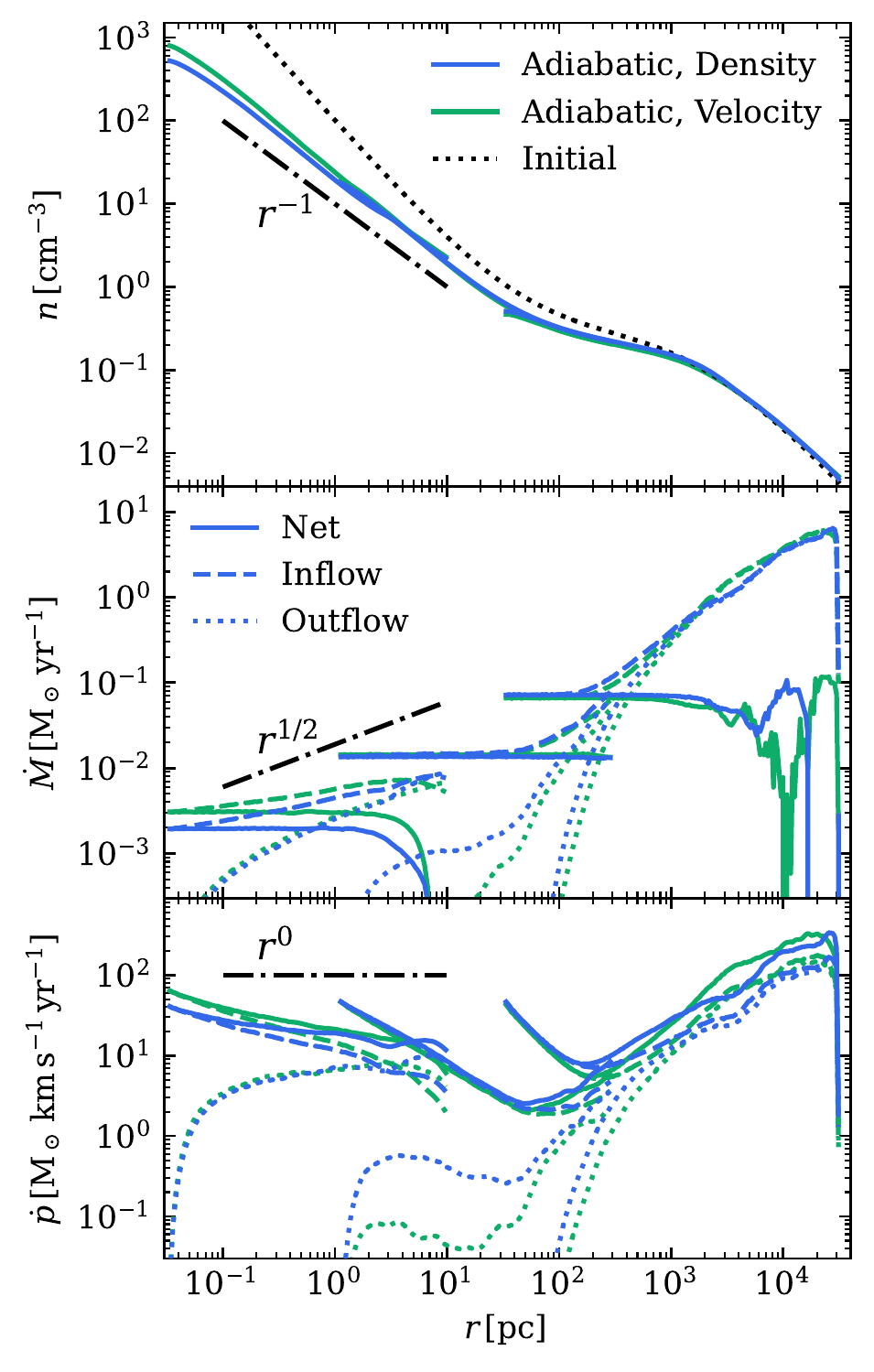}
    \caption{From top to bottom: radial profiles of time- and angle averaged- density, mass flux (dashed line for inflow and dotted line for outflow), and momentum flux for the adiabatic simulations with density (blue) and velocity (green) perturbations. We average between 100-200 Myr for $r>30\,\mathrm{pc}$, 1.5-2.5 Myr for $1<r<300\,\mathrm{pc}$, and 10-20 kyr for $r<10\,\mathrm{pc}$. The scalings are $\rho\propto r^{-1}$, $\dot{M}_\mathrm{in}\propto r^{1/2}$, and $\dot{p}_\mathrm{in}\propto r^{0}$. \label{fig:adb_radial}}
\end{figure}

\begin{figure}[ht!]
    \centering
    \includegraphics[width=\linewidth]{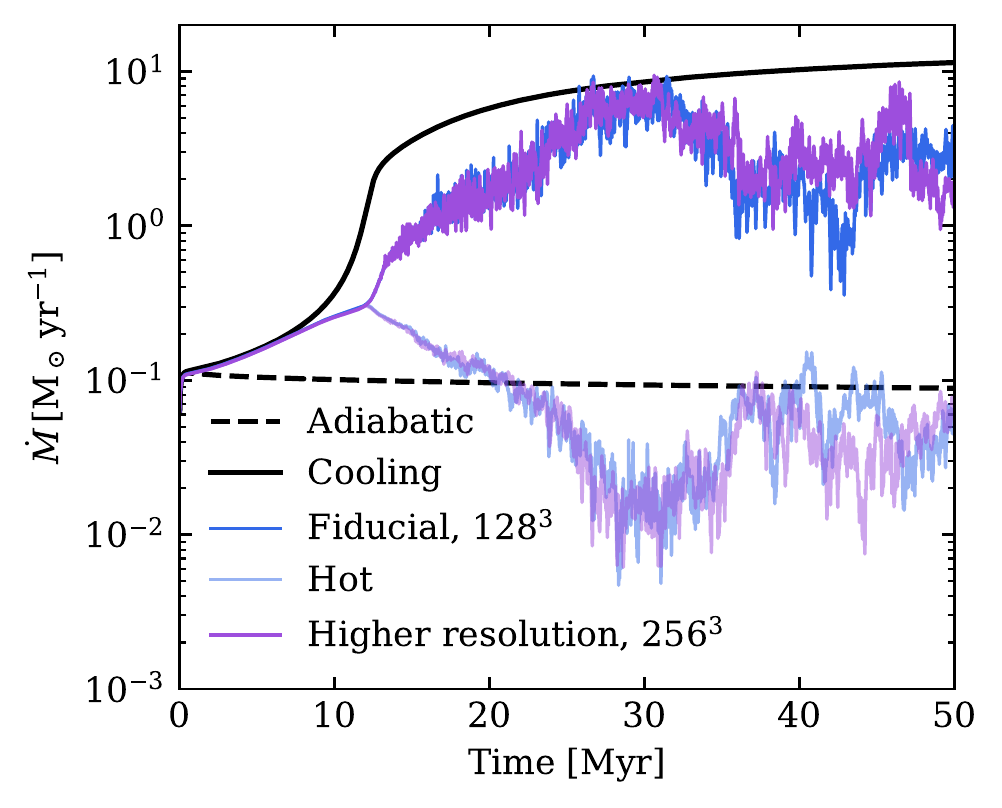}
    \caption{Convergence test by doubling the resolution (Model CHD10-r256). The accretion rate shows similar evolution to the fiducial simulation (Model CHD10) statistically, verifying a good convergence of the results at higher resolution. \label{fig:high_evo}}
\end{figure}

\begin{figure}[ht!]
    \centering
    \includegraphics[width=\linewidth]{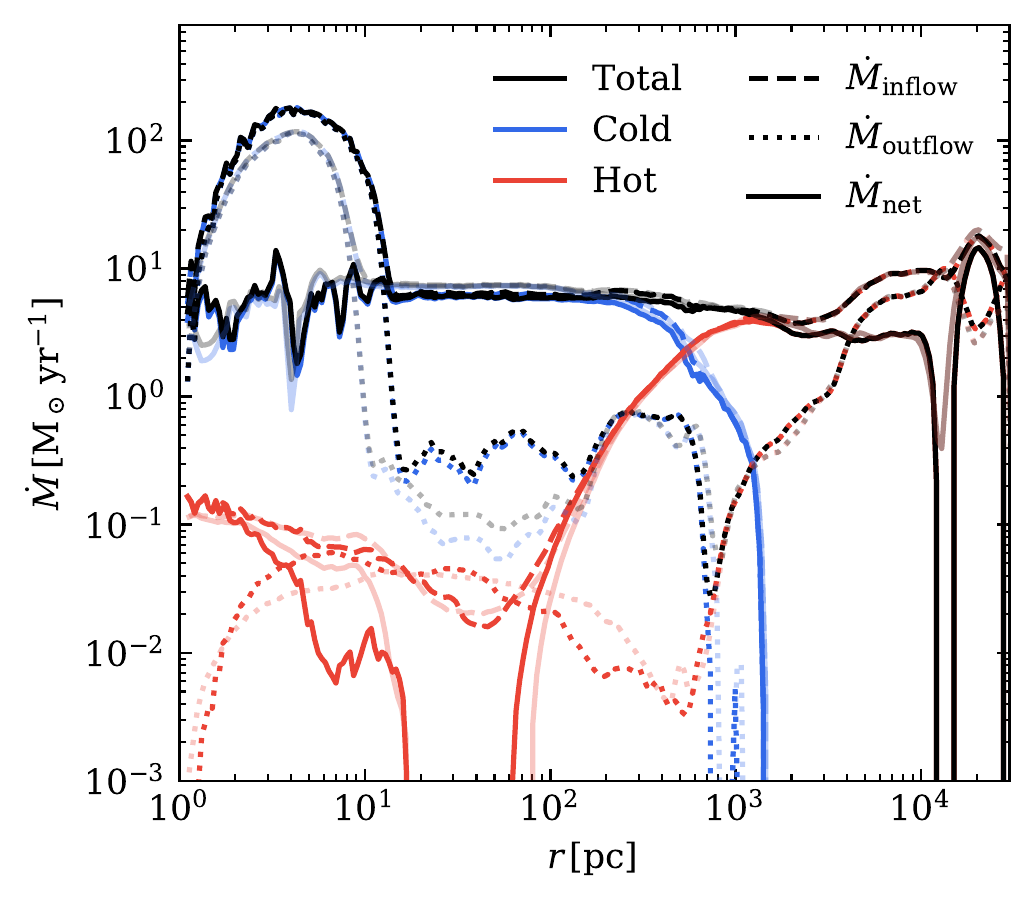}
    \caption{Radial profile of mass accretion rate for the chaotic stage of the simulation CHD12. The lighter lines show the simulation CHD15-t030 for comparison. The deviation between the two simulations is small.}
    \label{fig:one_radial}
\end{figure}

\begin{figure}[ht!]
    \centering
    \includegraphics[width=\linewidth]{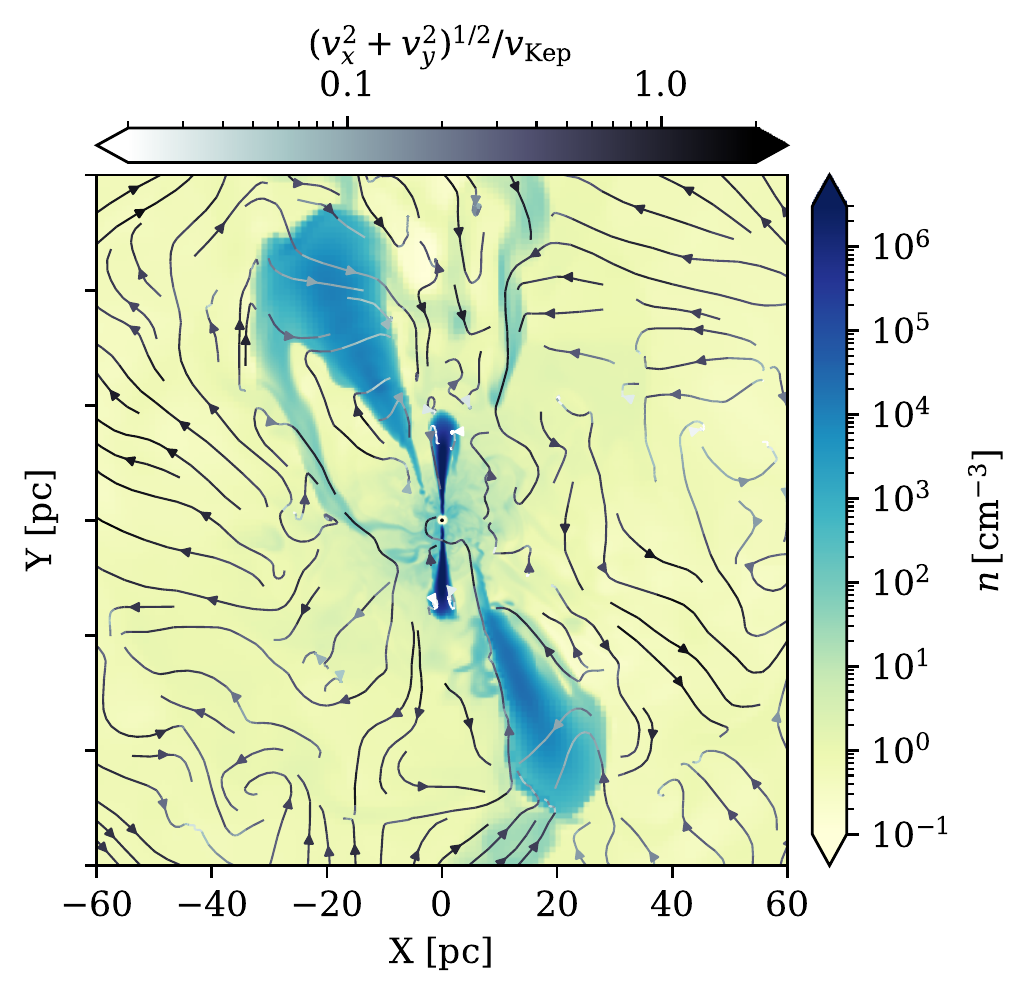}
    \caption{An example slice of density and streamlines of gas velocity through the $z = 0$ plane for the direct simulation with $r_\mathrm{in}=1\,\mathrm{pc}$ (Model CHD12). There is a two-tori structure. \label{fig:one_slice}}
\end{figure}

\section{Results} \label{sec:results}

In this section, we present the results of our fiducial suite of simulations, which include optically thin radiative cooling, heating, and turbulence. In general, we find there are two-phase (hot and cold) accretion flows. The cold accretion flows show two distinct stages that emerge from the simulations: a disky accretion flow and a chaotic accretion flow. Most of the time the disky accretion flow is a better description of the cold gas dynamics. \figu\ref{fig:fidu_render} shows a volume rendering and zoom-in of the central $\sim 1 \,\mathrm{kpc}$ region for both of these two stages.
The two snapshots are deliberately chosen to represent the typical structure of the chaotic (left) and disky (right) flows. We first describe in detail the properties of these accretion flows. Then we compare them with models with different levels of turbulence and idealized models with no cooling. Finally we present convergence tests.

Throughout the remainder of the paper, we define the gas with $T<T_\mathrm{cold}\equiv10 T_\mathrm{floor}$ as cold gas, gas with $T\ge T_\mathrm{hot}\equiv0.3 T_\mathrm{ini}(r)$ as hot gas (where $T_\mathrm{ini}(r)$ is the initial equilibrium solution), and gas in between ($T_\mathrm{cold}\le T< T_\mathrm{hot}$ ) as the intermediate gas.

\subsection{Fiducial Suite: with Radiative Cooling, Heating, and Density Perturbation }\label{subsec:fiducial}

In the fiducial suite of models, we include optically thin radiative cooling, radial shell averaged heating, and isobaric density perturbations (Models CHD10, CHD15..., and CHD20... in \tab\ref{tab:models}). 
With radiative cooling and heating producing approximate global thermal balance, the initial perturbations serve as the seed of local thermal instabilities, leading to the formation of a two-phase medium in about one cooling time. 

We first evolve the large scale simulations for $t_\mathrm{evo}=300\,\mathrm{Myr}$ with a inner radius of $r_\mathrm{in}=30\,\mathrm{pc}$. 
We pick several representative time points (30, 90, and 200 Myr, for CHD15-t030, CHD15-t090, and CHD15-t200, respectively) and restart the simulation with a smaller inner radius of $r_\mathrm{in}=1\,\mathrm{pc}$, and run the simulations for $t_\mathrm{evo}=2\,\mathrm{Myr}$ with $t_\mathrm{trans}=1\,\mathrm{Myr}$. As we shall describe below, the accretion flow at 90 Myr and 200 Myr is more disky, similar to a typical accretion disk, and accretion at 30 Myr is more chaotic without the presence of a torus or disk.
Then we again pick representative times for the middle scale simulations (Models CHD15...), restart them using an even smaller inner radius $r_\mathrm{in}=30\,\mathrm{mpc}\,(\approx100 r_\mathrm{g})$ and run them for $t_\mathrm{evo}=20\,\mathrm{kyr}$ with $t_\mathrm{trans}=10\,\mathrm{kyr}$ (Models CHD20...).

In \figus\ref{fig:fidu_t200_slice} and \ref{fig:fidu_t030_slice}, we present the two-dimensional $z=0$ slices of density, temperature, and the Bernoulli parameter $(E+P)/(\rho\Phi)-1$ with streamlines and their zoom-in from Models CHD20-t200 and CHD20-t030, respectively.
Generally, a cold dense torus or accretion disk (similar to that shown in \figus\ref{fig:fidu_t200_slice}) is formed during most of the time of our simulations.
Depending on the level of turbulence, the size and orientation of the disk change frequently due to the continuous inflow of filaments and clouds of cool gas. The cold torus resembles a typical cold disk in equilibrium. The size of the torus is typically $\sim 200$ pc, but can be larger or smaller, varying from tens to hundreds of parsec. At the same time, the orientation of the disk can differ greatly at different scales. 
The warps in the disk could be important for explaining the difference between the gas-based SMBH mass and the star/EHT-based mass in M87, which we will discuss in detail in \sect\ref{sec:disscussion}. 

Occasionally ($\lesssim 10\%$ of the time) the accretion flow is very chaotic and there is no disk structure at a larger scale ($10-1000\,\mathrm{pc}$). However, if we run a zoom-in simulation during those chaotic phases, we often find a tiny disk at smaller scales. This phase is shown in \figu\ref{fig:fidu_t030_slice}.

\subsubsection{Time Evolution}
The top panel of~\figu\ref{fig:fidu_evo} shows the time evolution of mass accretion rate of Model CHD10. After about one cooling time ($\sim 10\,\mathrm{Myr}$), the accretion becomes chaotic and stochastic everywhere, with the mass accretion rate through the inner boundary $r_\mathrm{in}=30\,\mathrm{pc}$ varying on time-scales of $\sim 1\,\mathrm{Myr}$ in a large range between the two simulations with spherically symmetric initial conditions: $\simeq 0.1\,M_\odot\,\mathrm{yr^{-1}}$ for adiabatic (Bondi) accretion (Model A10) and $\simeq 10\,M_\odot\,\mathrm{yr^{-1}}$ for accretion with cooling but no heating (Model C10), which is essentially set by mass cooling rate. 
The time-averaged accretion rate is approximately $1\,M_\odot\,\mathrm{yr^{-1}}$ during the simulation. The accretion rate contributed by the hot gas is around the Bondi rate, though it also varies modestly in time.

The mass-weighted angle-averaged value of the specific angular momentum,
\begin{equation}
    \vec{L}\equiv\frac{\int_S \rho\vec{r}\times\vec{v}\dd\Omega}{\int_S\rho\dd\Omega}, |L|\equiv\frac{\int_S \rho|\vec{r}\times\vec{v}|\dd\Omega}{\int_S\rho\dd\Omega},
\end{equation}
of the gas at the inner boundary (middle panel of \figu\ref{fig:fidu_evo}) is close to Keplerian most of the time due to the presence of a cold dense disk. 
However, the direction of angular momentum changes frequently due to the in-fall of clouds and filaments of cold gas. 
The typical time-scale for changing angular momentum completely is $10-100$ Myr, about $100-1000$ times the local free-fall time. 
This time-scale is comparable to or shorter than the viscous accretion time-scale, 
\begin{equation}
    \frac{t_\mathrm{vis}}{t_\mathrm{ff}}\sim \frac{1}{\alpha}\left(\frac{H}{r}\right)^{-2} = 10^4\left(\frac{\alpha}{0.01}\right)^{-1}\left(\frac{H}{0.1r}\right)^{-2},
\end{equation} 
suggesting that these flips in angular momentum might still be important even if we included MHD turbulence for angular momentum transport.

The mass and angular momentum are dominated by the cool disk in the inner region.
The mass of cold gas within $100\,\mathrm{pc}$ varies between $\sim 10^6\,M_\odot$ and $10^8\,M_\odot$ (bottom panel of \figu\ref{fig:fidu_evo}).
The contribution to the accretion rate from hot gas and cold gas varies in different cases, depending on the exact time and radius that we consider. In the long run, the time-averaged mass accretion rate may be set by the time-averaged rate at which cold gas is formed, $\dot{M}_\mathrm{cool}\simeq1\,M_\odot\,\mathrm{yr^{-1}}$ since all the cold gas would finally be accreted in the absence of star formation or feedback. But most of the time the accretion rate is smaller than the time-averaged value since the formation of the disk or torus suppresses the accretion rate.
The peaks of accretion rates close to the spherically symmetric cooling case ($\simeq10\,M_\odot\,\mathrm{yr^{-1}}$) are often associated with the rapid change of angular momentum of the cold gas.
Due to the cancellation of angular momentum, a considerable fraction of the cold gas can accrete to smaller radii. 

\figu\ref{fig:fidu_evo_15_20} plots the same quantities as in \figu\ref{fig:fidu_evo} but for Models CHD...-t030 and CHD...-t200. In the disk stage ($t=200\,\mathrm{Myr}$, red lines), when we trace the accretion flow down to a smaller region by gradually shrinking the inner radius, the accretion rate decreases significantly. The contribution from cold gas drops considerably and remains small during most of the simulation, though there are some peaks in the accretion rate associated with the infall of small cold clouds. 
The hot gas accretion rate also decreases, but essentially follows a power law $\dot{M}_\mathrm{hot} \propto r_\mathrm{in}^{0.5}$ (marked by the dotted line). We discuss the possible origin of this power-law in \sect\ref{sec:disscussion}. The angular momentum and mass around the inner boundary is still dominated by the cold gas (middle and bottom panels). The angular momentum of the gas through the inner boundary is similar to the Keplerian value due to the motion of the cold disk. However, the direction of angular momentum in the inner region can differ greatly from that in the outer region. 
In the chaotic stage ($t=30\,\mathrm{Myr}$, yellow lines), the accretion rate remains high at $\sim 1\,\mathrm{pc}$, but drops significantly at $\sim 0.03\,\mathrm{pc}$. This is associated with the formation of a small torus, as we will show below.

\subsubsection{Dynamics}

As is shown in \figu\ref{fig:fidu_t200_slice}, the gas appears in two distinct thermal phases: cold and hot phases with apparent boundaries. The cold gas is dense, remaining at a temperature similar to $T_\mathrm{floor}$, while the diffuse hot gas roughly keeps the virial temperature (the initial equilibrium temperature $T_\mathrm{ini}$). The cold gas forms a torus within $\sim 300\,\mathrm{pc}$. 
In \figu\ref{fig:fidu_t030_slice}, the cold gas is more chaotic, but we still find a small torus on parsec scales (nearly face-on by coincidence).
Generally, a small disk is formed easily in most cases, with sizes varying from tens of parsecs to hundreds of parsecs. When the disk is stable, the accretion rate tends to be low, similar to adiabatic accretion (Model A10). When some newly-formed cold clouds or filaments fall down to the central region, due to the cancellation of angular momentum, a considerable amount of gas flows into the sink cell. The accretion rate then increases significantly, approaching the spherically symmetric cooling case (Model C10). The direction of the disk in the inner region can be very different from the outer region. The hot flow is more chaotic at large scale but more laminar at small scale. 

In the suite of models with initial density perturbation, the initial angular momentum dispersion is very small. 
The flow tends to be very similar to the cold chaotic accretion model~\citet{Gaspari2013MNRAS.432.3401G} on larger scales. However, a large torus still forms intermittently. On small scales, a disk is present most of the time.

When there is a disk, the scale height of the cold disk is related to the temperature (sound speed) of the gas by
\begin{equation}
    \frac{H}{r}\simeq\frac{c_s}{v_\phi}\simeq 0.02\left(\frac{T_\mathrm{cold}}{2\times10^5\,\mathrm{K}}\right)^{1/2}\left(\frac{r}{1\,\mathrm{pc}}\right)^{1/2}.
\end{equation}
Thus it is extremely difficult to resolve the disk when the temperature is low and the radius is small. We set a temperature floor of $2\times 10^5\,\mathrm{K}$, higher than the physical value of $\simeq 10^4\,\mathrm{K}$, so as to resolve the structure of cold gas well within the simulation domain. 
A more physical cold disk would be denser and retain a smaller scale height by a factor of $\sim 5$. But the accretion rate and circularization of the cold gas should be similar, since they are basically determined by the angular momentum and inflow motion of the cold gas in the simulations, regardless of $T_\mathrm{floor}$. Going to small radii, the cold gas accretion is likely subdominant, so the impact of $T_\mathrm{floor}$ might be relatively small in this case. 
Within $\sim 0.3\,\mathrm{pc}$, the disk is aligned with the grid, which is perhaps due to the low resolution compared with the disk scale height. More generally, the formation of a torus is likely robust but the detailed thermodynamics, cooling, and heating properties of the disk might not be very reliable, especially on small scales. It remains to be investigated in the future about the exact effects of floors and resolutions on the accretion flow.

\subsubsection{Radial Profiles}

The time- and angle-averaged radial profiles of the gas and its two components (hot and cold) are shown in \figus\ref{fig:fidu_radial} and \ref{fig:fidu_radial_be_am}. 
Throughout this work, we define the time and angle average of a variable $A$ as
\begin{equation}
    \langle A \rangle \equiv \frac{\int_{t_0}^{t_1}\int_S A \dd\Omega \dd t}{(t_1-t_0)\int_S \dd\Omega} ,
\end{equation}
where $S$ is the area where we count the variable (the place of total, cold, and hot gas). The mass-weighted average is defined similarly by
\begin{equation}
    \langle A \rangle_m \equiv \frac{\langle A\rho \rangle}{\langle \rho \rangle}.
\end{equation}
Note that we do not directly obtain a global steady solution over the whole simulation domain shown in \figus\ref{fig:fidu_radial} and \ref{fig:fidu_radial_be_am}. This would require $10^4$ times more computational time, which is unrealistic in the foreseeable future. Instead, the solution at $r<10$ pc is essentially the steady solution with $r_\mathrm{in}=0.03\,\mathrm{pc}$ (CHD20-t200; which is run for 20 kyr), the solution at $10<r<1000$ pc is the solution with $r_\mathrm{in}=1\,\mathrm{pc}$ (CHD15-t200; which is run for 2 Myr), and the solution with $r>1000$ pc is the solution with $r_\mathrm{in}=30\,\mathrm{pc}$ (CHD10; which is run for 300 Myr).

As is shown in \figu\ref{fig:fidu_radial}, cold gas is generated within $\sim 2\,\mathrm{kpc}$. The temperature profiles of the hot gas follow a $T\propto r^{-1}$ scaling, as expected from conservation of energy, while the cold gas remains at the floor value of $\sim 2\times 10^{5}\,\mathrm{K}$. By contrast, the X-ray emission weighted temperature profile is rather flat (we use an emission range of $0.5 - 7\,\mathrm{keV}$). This implies that the gas that dominates $\sim$ keV emission is not hot gas at the viral temperature but rather the gas with T $\sim$ keV, over a wide range of radii. Note that the somewhat cooler gas with $T \sim \mathrm{keV} < T_\mathrm{vir}$ has higher density (at roughly constant pressure) which is why it dominates the keV luminosity. Observations of M87~\citep{Russell2015MNRAS.451..588R} suggested there are two significant temperature components within $2\,\mathrm{kpc}$ and did not find evidence for a temperature increase within $\sim 0.25\,\mathrm{kpc}$. The flat temperature profile is also reported in observations of the galaxy M84~(Bambic et al. 2022, in prep). This is consistent with our simulations. 

The angle-integrated mass inflow and outflow accretion rates are defined as
\begin{equation}
    \dot{M}_\mathrm{in}(r)= - \langle 4\pi\rho \cdot \min(v_r,0)r^2 \rangle ,
\end{equation}
and
\begin{equation}
    \dot{M}_\mathrm{out}(r)= \langle 4\pi\rho \cdot \max(v_r,0) r^2 \rangle.
\end{equation}
We also define the net accretion rate by $\dot{M}(r) \equiv \dot{M}_{\rm in}(r) - \dot{M}_{\rm out}(r)$.
For cold gas, both the gas inflow and outflow rates are large, but the net accretion rate through the inner boundary is relatively small.
For the hot gas, the inflow rate follows a $r^{1/2}$ scaling. However there is no strong hot outflow, so the net accretion rate also follows a radial dependence of $r^{1/2}$, rather than a constant.
This could imply an accumulation of hot gas but as is shown in \figu\ref{fig:fidu_evo}, the hot gas mass does not change. So the $\dot{M}\propto r^{1/2}$ scaling implies that the hot gas is mixed with the cold gas.

The density profile shows a scaling of $\rho\propto r^{-1}$ in \figu\ref{fig:fidu_radial}, which is different from the expected power law of $r^{-3/2}$ for $\gamma=5/3$ if the flow were adiabatic and spherically symmetric. This is consistent with scaling of hot gas accretion rate, since $\dot{M}\sim\rho vr^2\propto\rho r^{3/2}$. 
The cooling time in \figu\ref{fig:fidu_radial} shows that $t_\mathrm{cool}/t_\mathrm{ff}\equiv \langle E_\mathrm{int} \rangle/\langle Q_- \rangle/t_\mathrm{ff} > 10$ for hot gas at $1\,\mathrm{pc}\lesssim r\lesssim 1\,\mathrm{kpc}$. However, we note that the low-temperature tail of the hot gas can still cool efficiently due to radiative cooling and mixing with the cold gas. Actually, we find $t_\mathrm{cool}/t_\mathrm{ff} \sim 10$ for the intermediate gas in this region (though not shown here).

The Bernoulli parameter in \figu\ref{fig:fidu_radial_be_am} shows that, in contrast to the kpc scale, the hot gas is, on average, only slightly unbound inside $100 \,\mathrm{pc}$. As is shown by the snapshots in \figu\ref{fig:fidu_t200_slice}, part of hot gas is marginally bound within 1 pc. The cold gas is bound and dominated by rotational energy. 
So overall, the hot gas is unbound in most of the volume, but some hot gas is slightly bound within $\sim 1\,\mathrm{pc}$. The cold gas is completely bound with energy dominated by Keplerian rotation.

The radial profile of mass-weighted time- and angle-averaged absolute value of the specific angular momentum $|L|$ and the three components, $L_x$, $L_y$, and $L_z$, are shown in \figu\ref{fig:fidu_radial_be_am}. The hot gas is overall sub-Keplerian, while the cold gas within the torus possesses a Keplerian angular momentum on a small scale, but the direction varies. Due to the lack of viscosity, inefficient transport of angular momentum leads to a small accretion rate from cold gas. The accretion rate is mainly from the hot gas. It is unclear whether magnetic fields will change the relative hot and cold gas accretion significantly; we plan to investigate this in future work.

The radial dependence of net accretion rate shown in \figu\ref{fig:fidu_radial} is consistent with the cold gas mass accumulation in a large region shown in the bottom panel of \figu\ref{fig:fidu_evo}. 
This implies that we may not be in a statistically steady state. More time averaging may be needed, though the current averaging time is already $\gtrsim10^4$ times of the free-fall time at the inner radius. The accumulation of cold gas more likely reflects the fact that we do not model angular momentum transport in the cold disk, which likely precludes the existence of a global statistical steady state.

\subsubsection{Phase Distribution of Hot Gas and Cold Gas}
In analyzing the radial structure, we divided the gas into hot and cold phases. Here we discuss the mass distribution of the gas phases and the validity of the two-phase description.
In \figu\ref{fig:fidu_dist}, we show the mass distribution of density, temperature, and angular momentum for different radii.
There is clearly a bimodal distribution of density and temperature within $\sim 1\,\mathrm{kpc}$, indicating the two-phase structure of the gas.
The cold gas is generated at around 2 kpc and the two-phase medium extends to the central region. The mass of cold gas is estimated to be $\sim 10^6-10^8\,M_\odot$ in the whole simulation domain.
The hot gas has a virial temperature and density slightly lower than the initial solution.
Despite the bimodal distribution, there is a considerable amount of intermediate gas between  $T_\mathrm{cold}$ and $T_\mathrm{hot}$. This broad temperature range could be important in interpreting and modeling $\sim\,\mathrm{keV}$ X-ray emission (as indicated by the emission-weighted temperature in \figu\ref{fig:fidu_radial}).

A Keplerian cold disk is assembled around the black hole. The angular momentum distribution in \figu\ref{fig:fidu_dist} reflects the sub-Keplerian hot gas and nearly Keplerian cold gas. For the hot gas, the tail of the angular momentum distribution follows $\dd M/\dd\log_{10}L\propto L^2$ or equivalently, $\dd M/\dd L\propto L$.

\subsubsection{The Accretion Disk}

In the typical disky stage, there is generally a well-defined disk or torus structure at smaller radii. To analyze the properties of the disk, it is convenient to define a new coordinate system $(r,\theta',\phi')$ in which the $z'$-direction is aligned with the radial- and angle-averaged angular momentum of the cold gas.

We plot the $\phi'$-averaged density and temperature snapshots in \figus\ref{fig:phi_average} and the angular profiles in \ref{fig:angular}. The results indicate a cold dense disk structure at $\sim10-100\,\mathrm{pc}$ scale. 
The hot flow is laminar and falls down to the central sink region from the polar region or merges onto the cold disk by condensation. The cold disk closely follows Keplerian motion with a very small radial component. Most of the accretion is from hot gas in the polar region. The orientation of the disk changes within 10 pc so the inner disk is not described well in this coordinate system.

To investigate whether the cold disk would fragment into clumps and form stars due to spiral-mode gravitational instability, we calculate the Toomre parameter~\citep{Toomre1964ApJ...139.1217T}
\begin{equation}
    Q\equiv\frac{c_s\Omega}{\pi G \Sigma},
\end{equation}
where $\Sigma$ is the surface density. 
Given the temperature floor of $T_\mathrm{floor}=2\times10^5\,\mathrm{K}$ we use now, we find that the Toomre parameter of the disk reaches a minimum value of $\sim 20$ at $\sim100\,\mathrm{pc}$. 
For $T_\mathrm{floor}=10^4\,\mathrm{K}$, we would obtain $Q_\mathrm{min}\sim4$ if we suppose the surface density does not change, which is an appropriate assumption since the mass of the cold gas at $\sim100\,\mathrm{pc}$ is basically determined by the infall and accumulation of the cold gas, regardless of $T_\mathrm{floor}$.
In this case, the disk is marginally stable, but might be unstable if the surface density is higher or the temperature is even lower, which would lead to efficient fragmentation and star formation. In reality, in many cases the gas may be able to cool to below $10^4\,\mathrm{K}$, i.e., to become molecular $\sim 100\,\mathrm{K}$. This is observed in many ellipticals and groups and clusters~\citep{Davis2013Natur.494..328D,Davis2017MNRAS.468.4675D}, though probably not in M87. This implies that the gas may fragment and form stars in some cases and that star formation and supernovae feedback could be important to include.

\subsubsection{Chaotic Stage}
\label{sec:chaotic}
The chaotic stage is obtained only in relatively rare cases ($\lesssim10\%$ of the time) but is generally associated with a high accretion rate of cold gas. When a large amount of cold gas with negligible angular momentum is falling into the central region, the accretion flow is chaotic at smaller radii. The simulation at 30 Myr (Model CHD15-t030 and CHD20-t030) is a typical cold chaotic accretion flow, as is shown in \figus\ref{fig:fidu_t030_slice}, \ref{fig:fidu_evo}, and \ref{fig:fidu_evo_15_20}. Here we discuss the chaotic case in more detail. 

The radial profiles for the chaotic stage are shown in \figus\ref{fig:fidu_t030_radial} and \ref{fig:fidu_t030_radial_be_am}. The high accretion rate is sustained down to $\sim 3\,\mathrm{pc}$, where a small torus forms and the accretion rate decreases by one order of magnitude. 
The disk direction within $0.3\,\mathrm{pc}$ is different from the torus at $\sim 1\,\mathrm{pc}$. Similar to the disky stage, the disk is aligned with the grid at small scales, which is also possibly due to the low resolution compared with the disk scale height.
The hot gas is more chaotic compared with the disky stage, with a strong hot outflow on small scale. 

Due to the significant cooling, the heating rate around $10\,\mathrm{pc}$ is also slightly higher, leading to a relatively higher temperature and Bernoulli parameter for the hot gas, which is possibly associated with the strong hot outflow. The density of hot gas is also higher, possibly due to intense mixing with the cold clouds and filaments. The emission-weighted temperature profile is still relatively flat over a large range. On smaller scales ($\lesssim 1\,\mathrm{pc}$), a cold disk forms and the morphology and radial profiles tend to be similar to the disky cases.

\citet{Gaspari2013MNRAS.432.3401G} proposed that chaotic cold accretion could boost the accretion rate by up to two orders of magnitude. Here we find that while the accretion is overall chaotic and turbulent on large scales, a cold dense disk is formed easily in most of our simulations, especially on small scales ($\lesssim 100\,\mathrm{pc}$). Even for the very chaotic stage, there is usually a small disk within $\lesssim 3\,\mathrm{pc}$. The formation of the cold disk reduces the accretion rate toward the central region most of the time. We will discuss this in more detail in \sect\ref{sec:disscussion}.

\subsection{Accretion with Larger Angular Momentum}

Here we discuss the models in which we add velocity perturbations to the initial conditions (Models CHV... in \tab\ref{tab:models}). These perturbations imply a larger range in the specific angular momentum $|L|$, although the net angular momentum $\vec{L}$ remains small. 
The resulting flows are generally similar to the fiducial suite, except that the size of the disk tends to be larger, typically $\gtrsim 500\,\mathrm{pc}$, The orientation and size of the torus are changed by the inflow of new cold gas with different angular momentum, but this occurs infrequently. The typical time-scale for this process is around tens of million to hundreds of million years. 

The time evolution for Models CHV... is shown in \figu\ref{fig:vtur_evo}. The accretion around 100 Myr is suppressed because most of the cold gas is accumulated in the disk, contributing little to the accretion rate. The inflow of cold clouds leads to an increase in the accretion rate at later times.
Overall the accretion flows show similar behavior to the disk stage of the fiducial suite, except that the mass of cold gas is larger.
When we trace the accretion flow down to small scales, the accretion rate follows a similar radial scaling as the fiducial suite. Accretion is also dominated by the hot gas on small scales.
Although not shown here, we find that the flow morphology, radial profiles, and mass distribution show similar properties to the disky stage of the fiducial suite.

\subsection{Adiabatic Accretion with Turbulence}
In the fiducial suite of simulations, despite the presence of significant cooling, the cold disk does not contribute significantly to the accretion rate on small scales. Thus it is of interest to investigate the properties of adiabatic accretion with no cooling, so that cold gas is absent. Here we present adiabatic simulations with density and velocity perturbations (Models AD... and AV... in \tab\ref{tab:models}) and compare them with the fiducial suite that has cooling and heating and generates a multi-phase medium.

We plot the time evolution of mass accretion rate and angular momentum in \figu\ref{fig:adb_evo}. The accretion rate fluctuates, but overall is similar to the spherically symmetric case at $\sim 30\,\mathrm{pc}$. 
The absolute value of specific angular momentum through the inner boundary is sub-Keplerian and small compared with the simulations with cooling. The net angular momentum $\sqrt{L_x^2+L_y^2+L_z^2}$ is even smaller, implying a relatively large dispersion ($\sqrt{|L|^2-L_x^2-L_y^2-L_z^2}$).
As we restart the simulations with a smaller $r_\mathrm{in}$, the accretion rate follows the similar power law $r_\mathrm{in}^{1/2}$ as in the hot gas in the fiducial suite. 
Towards the inner region, the angular momentum also approaches Keplerian gradually, but it is still sub-Keplerian with a relatively large dispersion and never forms a Keplerian disk.
As is shown by the snapshots plotted in \figu\ref{fig:adb_slice}, though the fluctuations in the density and temperature are relatively small, the velocity field is highly turbulent. Most of the gas is unbound except at small scales. Despite many similarities with the fiducial suite, the streamlines in the adiabatic simulation are chaotic, remarkably different from the laminar streamlines in the disk stage of the fiducial suite.

We investigate the radial dependence of the accretion flow in the presence of moderate turbulence.
The radial profiles are shown in \figu\ref{fig:adb_radial}. We obtain a steady solution in $r<10$ pc with $r_\mathrm{in}=0.03\,\mathrm{pc}$, a steady solution in $1<r<300$ pc with $r_\mathrm{in}=1\,\mathrm{pc}$, and a steady solution in $30<r<10^4$ pc with $r_\mathrm{in}=30\,\mathrm{pc}$, as is shown clearly by the net accretion rate.
Adiabatic accretion also shows similar scaling to the fiducial cases. The density clearly scales with radius by $\rho\propto r^{-1}$. Combining the velocity scaling $v_r\propto r^{-1/2}$, this gives a power law of $\dot{M}_\mathrm{in}\propto r^{1/2}$ and constant momentum flux $\dot{p}_\mathrm{in}\propto r^{0}$. In contrast to the disk stage of the fiducial suite, however, there is a strong outflow of hot gas which leads to this scaling. Thus though both fiducial suite and adiabatic simulations show $\dot{M}_\mathrm{in}\propto r^{1/2}$, it seems to be for very different reasons. We return to this in \sect\ref{sec:disscussion}.

\subsection{Convergence Tests}

To examine the convergence with resolution, we present a higher-resolution simulation with a root grid of $256^3$ (Model CHD10-r256 in \tab\ref{tab:models}). In this case, the resolution is doubled everywhere compared with the fiducial simulation (Model CHD10). As is shown in~\figu\ref{fig:high_evo}, the history of the accretion rate is very similar to the fiducial case. The morphology, dynamics, and radial profiles of the two simulations are also very similar. This further confirms the validity of our results. 

To test whether our method of changing the inner boundary affects the outcome, we directly simulate the accretion flow with $r_\mathrm{in}=1\,\mathrm{pc}$ using 12 levels of mesh refinement with the finest resolution of $\Delta x=0.12\,\mathrm{pc}$ for a long evolution time of $\sim 100\,\mathrm{Myr}$ (Model CHD12 in \tab\ref{tab:models}). 
This simulation required nearly one million core hours due to the extremely small time step. 
The results essentially confirm our results from the fiducial suite of simulations. We find that most of the time, the accretion rate at small scales stays low compared with the case with a larger $r_\mathrm{in}$. Occasionally the accretion rate is higher than the large-scale simulation due to the accumulation of cold gas. The hot gas accretion rate is also systematically lower than the fiducial case CHD10. In \figu\ref{fig:one_radial}, we plot the radial profile of accretion rate in the chaotic stage and compare it with the simulation CHD15-t030. Across the range of radii, the accretion rate only shows slight deviations from the model CHD15-t030. This indicates that the changing inner boundary strategy does not introduce significant systematic deviations at smaller radii.

Interestingly, we find a \textit{two-tori} structure during the late time of this simulation, as shown by the $z=0$ slice of density with streamlines in \figu\ref{fig:one_slice}. 
Before the old torus is depleted, a new torus pointing to a different direction forms on a larger scale due to the infalling of cold clouds with different angular momentum from the old torus. 
The size of the old torus is $\sim 20\,\mathrm{pc}$ and the new torus is $\sim 60\,\mathrm{pc}$. This structure sustained for a long time ($\sim 30\,\mathrm{Myr}$). Similar to the warped disk found in the fiducial suite, this also implies that the direction of the accretion disk could differ greatly on different scales, though presumably for various reasons. 
The potential caveat is the grid alignment of inner torus, which is possibly due to the low resolution ($r_\mathrm{in}/\Delta x_\mathrm{min}\approx8$). 
Given that each torus is a massive structure, the inclusion of self-gravity may change the evolution of each torus and their interaction qualitatively.

\begin{deluxetable*}{lrrrrrrrrr}
    \tablenum{2}
    \tablecaption{Time-averaged mass accretion rates \label{tab:mdots}}
    \tablewidth{30pt}
    \tablehead{
    \colhead{Model} & \colhead{Inner} & \colhead{Time} & \colhead{Accretion} & \colhead{$f_{\dot{M}}$} & \colhead{$f_{\dot{M}}$} & \colhead{$f_{\dot{M}}$} & \colhead{$f_t$} & \colhead{$f_t$} & \colhead{$f_t$}\\[-0.2cm]
    \colhead{Label} & \colhead{Radius} & \colhead{Range} & \colhead{Rate} & \colhead{Cold} & \colhead{Int.} & \colhead{Hot} & \colhead{Cold} & \colhead{Int.} & \colhead{Hot}\\[-0.8cm]
    }
    \decimalcolnumbers
    \startdata
    CHD10 & 30 pc & 20-300 Myr & $1.5\,M_\odot\,\mathrm{yr^{-1}}$ & 93\% & 2.1\% & 4.9\% & 97\% & 0.0\% & 3.4\%\\
    CHD10 & 30 pc & 200-203 Myr & $0.80\,M_\odot\,\mathrm{yr^{-1}}$ & 90\% & 1.7\% & 8.6\% & 98\% & 0.0\% & 1.7\%\\
    CHD15-t200 & 1.0 pc & 2-3 Myr & $0.034\,M_\odot\,\mathrm{yr^{-1}}$ & 36\% & 24\% & 40\% & 35\% & 7.9\% & 57\%\\
    CHD20-t200 & 0.03 pc & 20-30 kyr & $0.010\,M_\odot\,\mathrm{yr^{-1}}$ & 0.1\% & 28\% & 72\% & 0.0\% & 5.2\% & 95\%\\
    \enddata
    \tablecomments{The mass accretion rate and the role of different phases of gas for Models CHD10 and CHD..-t200. Here $f_{\dot{M}}$ is the mass fraction contributed by the corresponding gas phase and $f_{t}$ is the time fraction that the corresponding phase dominates the accretion.}
\end{deluxetable*}

\begin{figure}[ht!]
    \centering
    \includegraphics[width=\linewidth]{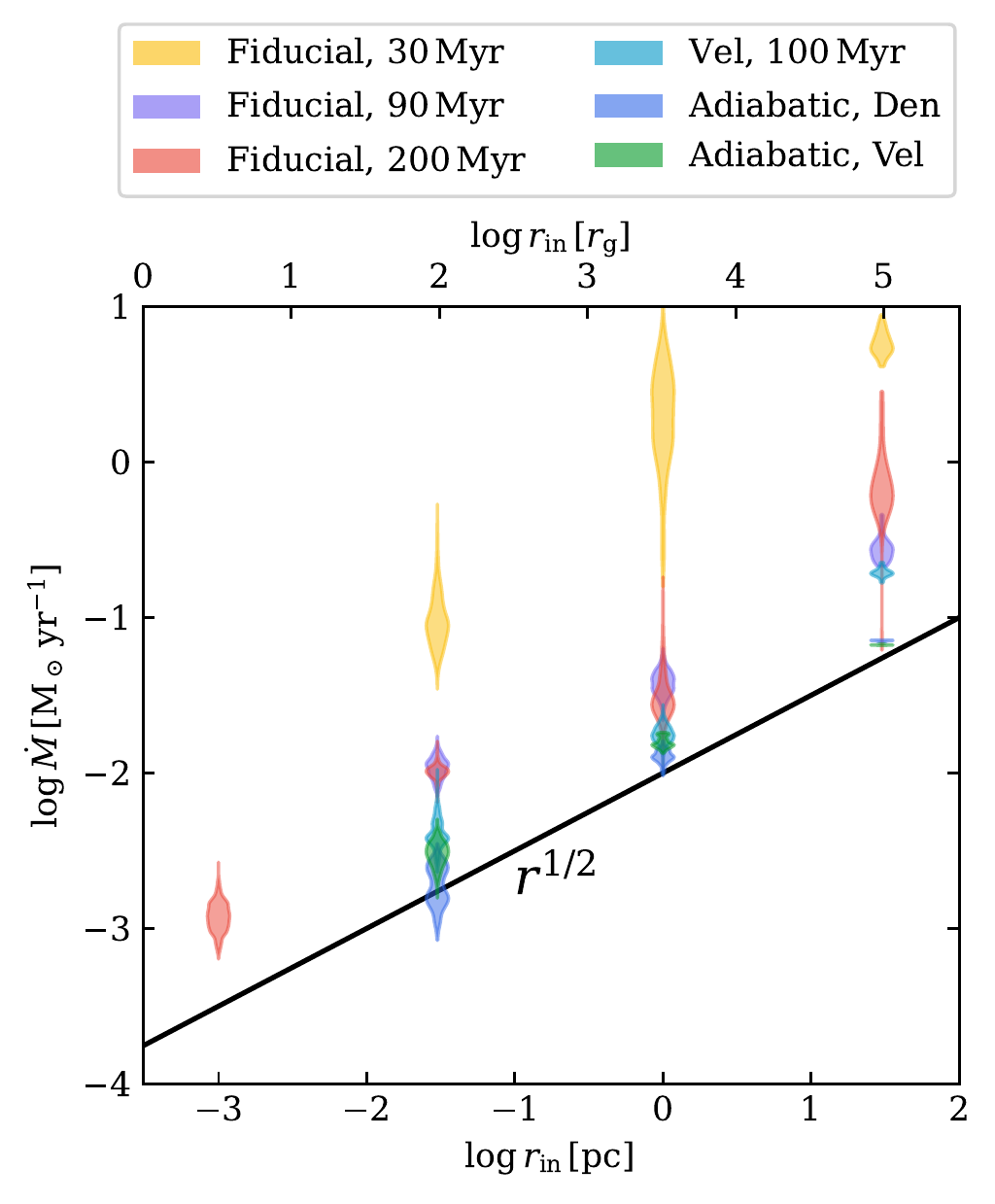}
    \caption{Relationship between mass accretion rate and inner radius for different runs. The accretion rates are values measured with a time range of $t=3\,\mathrm{Myr}$ for $r_\mathrm{in}=30\,\mathrm{pc}$, $t=1\,\mathrm{Myr}$ for $r_\mathrm{in}=1\,\mathrm{pc}$, $t=10\,\mathrm{kyr}$ for $r_\mathrm{in}=30\,\mathrm{mpc}$, and $t=100\,\mathrm{yr}$ for $r_\mathrm{in}=1\,\mathrm{mpc}\,(3r_\mathrm{g})$. The simulations with the highest accretion rate (Fiducial, 30 Myr; yellow) are in the cold chaotic accretion phase. Overall, we find a universal power law of $\dot{M}\propto r_\mathrm{in}^{1/2}$ over a large dynamic range from Bondi scale to horizon scale, spanning three to five orders of magnitude. The normalization of the accretion rate vs. radius is similar for simulations {\em not} in the cold, chaotic accretion phase. \label{fig:powerlaw}}
\end{figure}

\section{Discussion} \label{sec:disscussion}

\subsection{Towards the Event Horizon}
\label{subsec:scaling}

In \figu\ref{fig:powerlaw}, we summarize the accretion rate through the inner boundary at different times in Model CHD as well as Models CHV, AD, and AV. In \tab\ref{tab:mdots}, we list the time-averaged mass accretion rate and the contribution from different phases of gas for Model CHD10 and CHD...-t200 (the typical disk stage). The radial dependence in \figu\ref{fig:powerlaw} is well described by the $r_\mathrm{in}^{1/2}$ scaling, especially when cold inflow is negligible (i.e., except for the yellow points for Model CHD at t=30Myr).
The sink region boundary condition in our simulations means the absence of radial pressure support, leading to a negligible outflow rate through the inner boundary and increasing the net accretion rate. This effect is reasonable only if $r_\mathrm{in}$ is the event horizon of the black hole. Unfortunately, covering such a large physical region is a formidable challenge due to the small time-steps required.

Fortunately, we can reasonably extrapolate the accretion rate down to smaller radii based on the self-similar radial dependence of the accretion flow in the simulations we have conducted.
The effect of the inner boundary is to force $v_r(r_\mathrm{in})\approx-c_\mathrm{in}\propto r_\mathrm{in}^{-1/2}$. Combining $\rho\propto r^{-1}$, we expect $\dot{M}\propto r^{1/2}$.
As is shown in \figu\ref{fig:powerlaw}, we find a mass accretion rate of $10^{-1}-1\,M_\odot\,\mathrm{yr^{-1}}$ at 30 pc, $10^{-2}-10^{-1}\,M_\odot\,\mathrm{yr^{-1}}$ at 1 pc, and $10^{-3}-10^{-2}\,M_\odot\,\mathrm{yr^{-1}}$ at 0.03 pc. The exception is the chaotic case (fiducial simulation at 30 Myr, Models CHD...-t030), which is relatively rare, and which produces a larger accretion rate at small radii. By extrapolating our simulation results down to the event horizon, we obtain an accretion rate $\sim10^{-4}-10^{-3}\,M_\odot\,\mathrm{yr^{-1}}$. 
We find a fairly good agreement in the simulations.
The accretion rate is a thousand times smaller than the Bondi accretion rate prediction. It is remarkably consistent with $\dot{M}\sim(3-20)\times10^\mathrm{-4}\,M_\odot\,\mathrm{yr^{-1}}$ from EHT observations~\citep{M87EHT_VIII_2021ApJ...910L..13E}.

As a test of this scaling of the accretion rate with inner boundary radius, we furthermore ran one simulation with $r_\mathrm{in}=1\,\mathrm{mpc}\,(3r_\mathrm{g}!)$ and 25 levels of mesh refinement (Model CHD25-t200 in \tab\ref{tab:models}) to explore the accretion around the event horizon. This result is shown in \figu\ref{fig:powerlaw}, and yields a accretion rate $\sim 10^{-3}\,M_\odot\,\mathrm{yr^{-1}}$, consistent with the extrapolation. We note that in this case, the resolution is insufficient to capture the dynamics of the cold disk and we neglect crucial physics (general relativity, magnetic field, radiation, collisionless effects, etc) that we plan to include in the future. We also do not aim to give a precise prediction of the accretion rate around the event horizon using this simulation. But nonetheless, it demonstrates that the scaling $\dot M \sim \dot M_{\rm Bondi} (r_\mathrm{g}/r_\mathrm{B})^{1/2}$ might be a good prescription for the accretion rate on horizon scales.

The approach used in this paper, when applied to future MHD simulations which we plan to run,  will enable us to perform MHD simulations over the entire dynamic range of the accretion flow. This will have the potential to benefit more realistic initial or boundary conditions for GRMHD simulations on horizon scales, instead of an idealized torus with specified angular momentum.

\subsection{Physics of Hot Gas Accretion}
It remains unclear how to self-consistently link the accretion rate and dynamics at the Bondi scale to the event horizon for hot gas accretion. For adiabatic accretion flows, there exist three possible solutions assuming self-similarity~\citep{Gruzinov2013arXiv1311.5813G}: 1. the spherically symmetric Bondi flow, with $\rho\propto r^{-3/2}$ and constant mass flux; 2. the Convection Dominated Accretion Flow (CDAF) with $\rho\propto r^{-1/2}$, characterized by strong entropy inversion and constant energy flux with $\dot{M}\propto r$; 3. a turbulent accretion flow identified by constant momentum flux with $\rho\propto r^{-1}$ and $\dot{M}\propto r^{1/2}$.

As is discussed in \sect\ref{subsec:scaling}, the third scaling of $\dot{M}\propto r^{1/2}$ is an excellent description of the accretion flow in our simulations in the presence of moderate turbulence. It is worth noting that there is some evidence that this scaling is universal if the accretion is nearly adiabatic and modestly turbulent. Similar scaling is reported in various contexts and with varying physics, including some MHD and GRMHD simulations of radiatively inefficient black hole accretion~\citep{Pang2011MNRAS.415.1228P,White2020ApJ...891...63W}, fueling of Sagittarius A* via the stellar winds of Wolf-Rayet stars within the central parsec of the galactic center~\citep{Ressler2018MNRAS.478.3544R, Ressler2020ApJ...896L...6R}, Bondi-Hoyle-Lyttleton accretion in supergiant X-ray binaries~\citep{Xu2019MNRAS.488.5162X}, wandering massive black holes in the outskirts of galaxies~\citep{Guo2020ApJ...901...39G}, and adiabatic super-Eddington accretion~\citep{Hu2022ApJ...934..132H}.

It is somewhat surprising that our simulations with multiphase gas, heating, and cooling produce a similar accretion rate $\propto r_\mathrm{in}^{1/2}$ to our adiabatic simulations of Bondi accretion with turbulence. The reason this is surprising is that the hot gas dynamics is very different in these two cases. 
There are no strong outflows of hot gas when cooling produces a significant cold disk. Instead, we find mixing of hot and cool gas that depletes the amount of hot gas at smaller radii. The streamlines of the hot gas with cooling (\figu\ref{fig:fidu_t200_slice}) and in the adiabatic simulations (\figu\ref{fig:adb_slice}) highlight the large differences in the hot gas physics: in the former case the hot gas settles relatively smoothly onto the cold disk while in the latter case the hot gas is much more turbulent. 
The fundamental explanation in most of the literature for a reduction in the accretion rate relative to the Bondi rate ($\dot{M} \propto r^p$ for some $1 > p > 0$) is that outflows suppress the accretion rate at small radii \citep{Blandford1999}, with some other conserved quantity setting the density profile and thus $\dot{M}(r)$. This is essentially the argument for $\dot{M}\propto r^{1/2}$ or $\dot{M}\propto r$ in the previous literature \citep[e.g.,][]{Narayan2000ApJ...539..798N,Gruzinov2013arXiv1311.5813G}. However, in our simulations with multiphase gas, $\dot{M}$ is suppressed but \textit{not} for this reason. Given this different physics, we do not see any particularly compelling reason to expect $\dot{M}(r)$ to be similar in our simulations with and without cooling, but yet they seem to be; this is an interesting puzzle.

\subsection{Physics of Cold Gas Accretion}
In our simulations, cold gas accretion includes two distinct stages: the typical disk stage and the relatively rare chaotic stage. We find that while the accretion is overall chaotic and turbulent, there is always a cold dense disk or torus in the inner hundreds of parsec (or, in the most extreme cases, in the inner few pc). However, when it is present the cold disk does not generally contribute significantly to the accretion rate at small radii, which is dominated by the hot gas. This is partially by construction, since we do not include magnetic fields or viscosity in our simulations. As a result, the accretion rate of cold gas typically decreases with decreasing inner radius because the cold gas accumulates in a disk.

Consistent with previous results~\citep{Li&Bryan2012ApJ...747...26L,Gaspari2013MNRAS.432.3401G}, we find that there can occasionally be a significant accretion rate of cold gas at intermediate radii. 
The driver of this accretion is the large fluctuations in the angular momentum of the cold gas at large radii and the resulting cancellation of angular momentum via frequent collisions between the inflowing cold gas and the disk. 
A series of analytical calculations~\citep{King&Pringle2006MNRAS.373L..90K, King&Pringle2007MNRAS.377L..25K} postulated that BHs may grow through a series of randomly oriented accretion events analogous to what we find in our simulations. 
It is possible that the random fluctuations in inflowing angular momentum that promote occasional cold gas accretion may still be important even in the presence of turbulence and magnetic fields in the cold gas disk. This is because the viscous time in a thin disk is relatively long compared to the time over which the inflowing angular momentum changes in our simulations. On the other hand, large scale magnetic stresses can remove angular momentum from the inflowing cold gas much more efficiently than local turbulence; we will study the role of these large-scale magnetic stress in future work.

One important feature of the cold disk in our simulations is that the orientation of the disk can differ significantly on different scales. Furthermore, we find the co-existence of two tori with different sizes and orientations in one simulation covering a large radial dynamic range (Model CHD12 in \tab\ref{tab:models}).
The warps that we find could be important for explaining the difference between the gas-based SMBH mass in M87 of $3\times10^9\,M_\odot$ and the star/EHT-based mass of $6\times10^9\,M_\odot$. The reason is that the gas-based mass inference is sensitive to inclination (e.g., \citealt{Walsh2013ApJ...770...86W}). If there is a warp and the inclination varies on the scale of the observed disk, that could bias the mass inferences from gas kinematics. 

\subsection{Accretion of Hot vs. Cold Gas}

AGN feedback is believed to play a key role in galaxy evolution and is crucial for cosmological simulations to produce a realistic population of massive galaxies, groups, and galaxy clusters~\citep{Fabian2012ARA&A..50..455F,Heckman2014ARA&A..52..589H,Somerville2015ARA&A..53...51S}. Our work has the potential to provide a more physical and accurate subgrid model of AGN feedback in groups and clusters by linking the accretion rate at a larger scale to the accretion rate on horizon scales. 
In cosmological simulations the Bondi accretion rate is typically utilized to model how the black hole accretes gas from its surroundings~\citep{Somerville2015ARA&A..53...51S,Pillepich2018MNRAS.473.4077P}. By contrast, accretion of cold gas is the key in current high-resolution simulations of how AGN feedback balances cooling in clusters ~\citep[][etc]{Li&Bryan2012ApJ...747...26L,Gaspari2013MNRAS.432.3401G}. 
The latter models are essentially a limit cycle in which cold gas accretion generates feedback that suppresses cooling for a period of time (a few cooling times of hot gas), then a new burst of cold gas accretion sets in. 
We find, by contrast, that the accretion at small radii is dominated by the hot gas most of the time (see Table \ref{tab:mdots}). In future work it will be critical to understand how the interplay between hot gas and cold gas accretion changes with the inclusion of magnetic fields, which can further drive accretion of the cold gas. This will clarify whether cold or hot gas dominates the time-averaged accretion and feedback in massive galaxies. Thermal conduction and viscosity may also be important because they can alter the dynamics of the hot gas and the mixing between cool and hot phases that is important in setting the hot gas accretion rate in our simulations~\citep{Nayakshin2004astro.ph..2469N}. 

\subsection{Comparison to Previous Work}
The results presented here on the interplay between heating, cooling, and accretion in massive galaxies and galaxy clusters are closely related to previous simulations by \citet{Li&Bryan2012ApJ...747...26L} and \citet{Gaspari2013MNRAS.432.3401G} (among others).
Relative to these earlier works, our main improvements include:
\begin{itemize}
    \item We focus on connecting physics on multiple scales, adopting a very small inner boundary radius, generally spanning six orders of magnitude down to 0.03 pc, with one test case down to $\sim 3 r_\mathrm{g}$. 
    \item We achieve a high resolution even around the sink region, which is crucial in resolving the cold gas.
    \item We run the simulation for a very long time to investigate the statistically steady state.
\end{itemize}

\citet{Li&Bryan2012ApJ...747...26L} studied the accretion of gas with modest rotation from Mpc scales down to pc scales. They found that the cooling gas rapidly forms a thin disk, as we do, but did not trace the evolution of the disk for a long time.
\citet{Gaspari2013MNRAS.432.3401G} carried out simulations linking the physics at tens of kpc in clusters to sub-pc scales.
Overall, they found that chaotic cold accretion was more prominent/important than we find in our simulations. Part of this could be due to the relatively short duration of their simulations, which was $\sim40\,\mathrm{Myr}$, only a few $t_\mathrm{cool}$. We find that chaotic accretion is more prominent at early times comparable to these in our simulations, while the disk phase dominates later as the angular momentum of the cold gas continues to accumulate.
Another difference is that we only include turbulence (which generates angular momentum fluctuations) in the initial conditions. \citet{Gaspari2013MNRAS.432.3401G} adopted continuous turbulence driving, which might promote chaotic accretion by changing the angular momentum of the inflowing gas. Simulations with more realistic turbulence driven by AGN, substructure, and/or stellar feedback would be valuable to better understand the cold gas dynamics on small scales in massive galaxies.

\section{Summary} \label{sec:summary}
We have presented a series of high-resolution, three-dimensional hydrodynamic simulations of the fueling of SMBHs in elliptical galaxies, taking M87 as a typical case. The key physical ingredients are radiative cooling and a phenomenological heating model. The simulations span more than 6 orders of magnitude in radius, reaching all the way down to sub-parsec scales with one test simulation reaching to near the event horizon.

Our main conclusions are: 
\begin{itemize}
    \item The accretion flow takes the form of multiphase gas at radii less than about a kpc. Most of the time accretion at the smallest radii is dominated by hot virialized gas. Nonetheless, simple adiabatic simulations of hot gas accretion do not reproduce the dynamics found in our full simulations: the hot gas inflow rate in the latter is strongly influenced by condensation onto the cold disk formed by radiative cooling.
    \item Cold gas accretion includes two dynamically distinct stages: the typical disk stage in which the cold gas resides in a rotationally supported disk and relatively rare chaotic stages in which the disk is broken up and the cold gas inflows via chaotic streams. Disks are formed easily in most of our simulations, especially on small scales. Chaotic cold accretion is obtained only in a fine-tuned parameter space. 
    \item The mass accretion rate scales with radius as $\dot{M}_\mathrm{net}\propto r_\mathrm{in}^{1/2}$ when hot gas dominates. By extrapolating the accretion rate down to the event horizon, we obtain $\dot{M}\simeq10^\mathrm{-4}-10^\mathrm{-3}\,M_\odot\,\mathrm{yr^{-1}}$, similar to $(3-20)\times10^\mathrm{-4}\,M_\odot\,\mathrm{yr^{-1}}$ from EHT observations. The same scaling of $\dot M \propto r^{1/2}$ can explain both observations and simulations of the Galactic Center \citep{Ressler2020ApJ...896L...6R}. This prescription $\dot{M} \propto r^{1/2}$ is thus an attractive option for connecting hot gas properties at large radii (in observations or cosmological simulations) to the accretion rate near the event horizon. We suggest that this is a more physical and accurate subgrid model of SMBH fueling by hot gas than the standard Bondi accretion rate assumption typically used in large-scale cosmological simulations.
    \item The relative contribution of hot and cold gas to the total accretion rate varies on different spatial scales and on different time-scales. Most of the time, accretion at small radii is primarily due to hot gas, and only in rare periods does cold gas dominate. It is possible, however, that cold gas accretion could still dominate the time-averaged accretion rate, but simulations with more physics in the cold gas (e.g., magnetic fields) are needed to quantitatively determine this. 
    \item The orientation of the gas disk can differ significantly on different spatial scales, primarily due to the inflow of gas with a wide range of angular momentum. Such warps in the disk could be important for explaining the difference between the gas-based SMBH mass in M87 of $3\times10^9\,M_\odot$ and the star/EHT-based mass of $6\times10^9\,M_\odot$ (because the gas-based SMBH mass is sensitive to the inclination of the gas disk on scales that are not well resolved observationally, and thus to the presence of warps).
\end{itemize}

Because of insufficient resolution and the absence of crucial physics (e.g., magnetic fields), our results on the internal dynamics of the cold gas are likely less reliable than our results on the dynamics of the hot gas; inclusion of anisotropic conduction and viscosity would further bolster the realism of our treatment of the hot gas. 
MHD simulations with these more realistic physics ingredients will be carried out in the near future. Using these results, we can also provide more realistic initial and boundary conditions for GRMHD simulations of accretion on event horizon scales.

\section*{Acknowledgement}
MG would like to thank Patrick Mullen, Amy Secunda, Eve Ostriker, and Philip Hopkins for helpful discussions during the development of this work. 
This work was supported by a grant from the Simons Foundation (888968, E.C. Ostriker, Princeton University PI) as part of the Learning the Universe Collaboration.
JS acknowledges support from NASA grants HST-GO-15890.023A and 80NSSC21K0496, and from the Eric and Wendy Schmidt Fund for Strategic Innovation. EQ was supported in part by a Simons Investigator grant from the Simons Foundation and NSF AST grant 2107872. 
The authors are pleased to acknowledge that the work reported on in this paper was substantially performed using the Princeton Research Computing resources at Princeton University which is consortium of groups led by the Princeton Institute for Computational Science and Engineering (PICSciE) and Office of Information Technology's Research Computing.

\vspace{5mm}

\software{\athena\ \citep{Stone2020ApJS..249....4S},
          }

\appendix

\section{Reference Simulations}
As tests of our model, here we present the results of the reference simulations (Models A13, C13, CD, and CV listed in \tab\ref{tab:models}).

\begin{figure*}[ht!]
    \centering
    \includegraphics[width=0.9\linewidth]{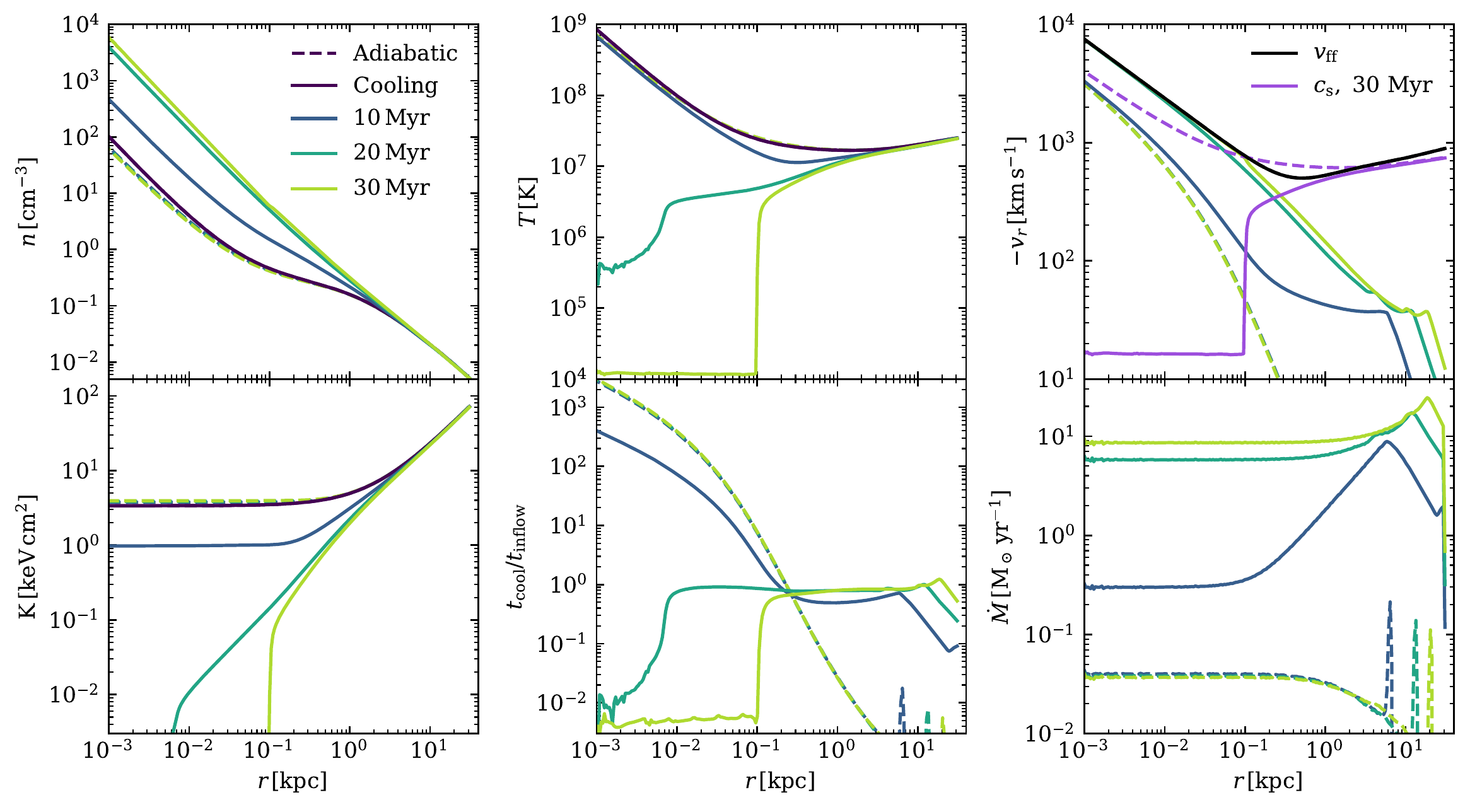}
    \caption{Radial profiles of angle-averaged density (top left), temperature (top middle left), velocity (top left), entropy (bottom left), cooling time (bottom middle), and accretion rate (bottom right) for the spherically symmetric adiabatic (dashed) and cooling (solid) simulations. The adiabatic accretion basically follows the Bondi solution and the cold accretion forms a cooling accretion flow, boosting the accretion by a factor of $\sim 100$.\label{fig:sph_radial}}
\end{figure*}

\begin{figure}[ht!]
    \centering
    \includegraphics[width=\linewidth]{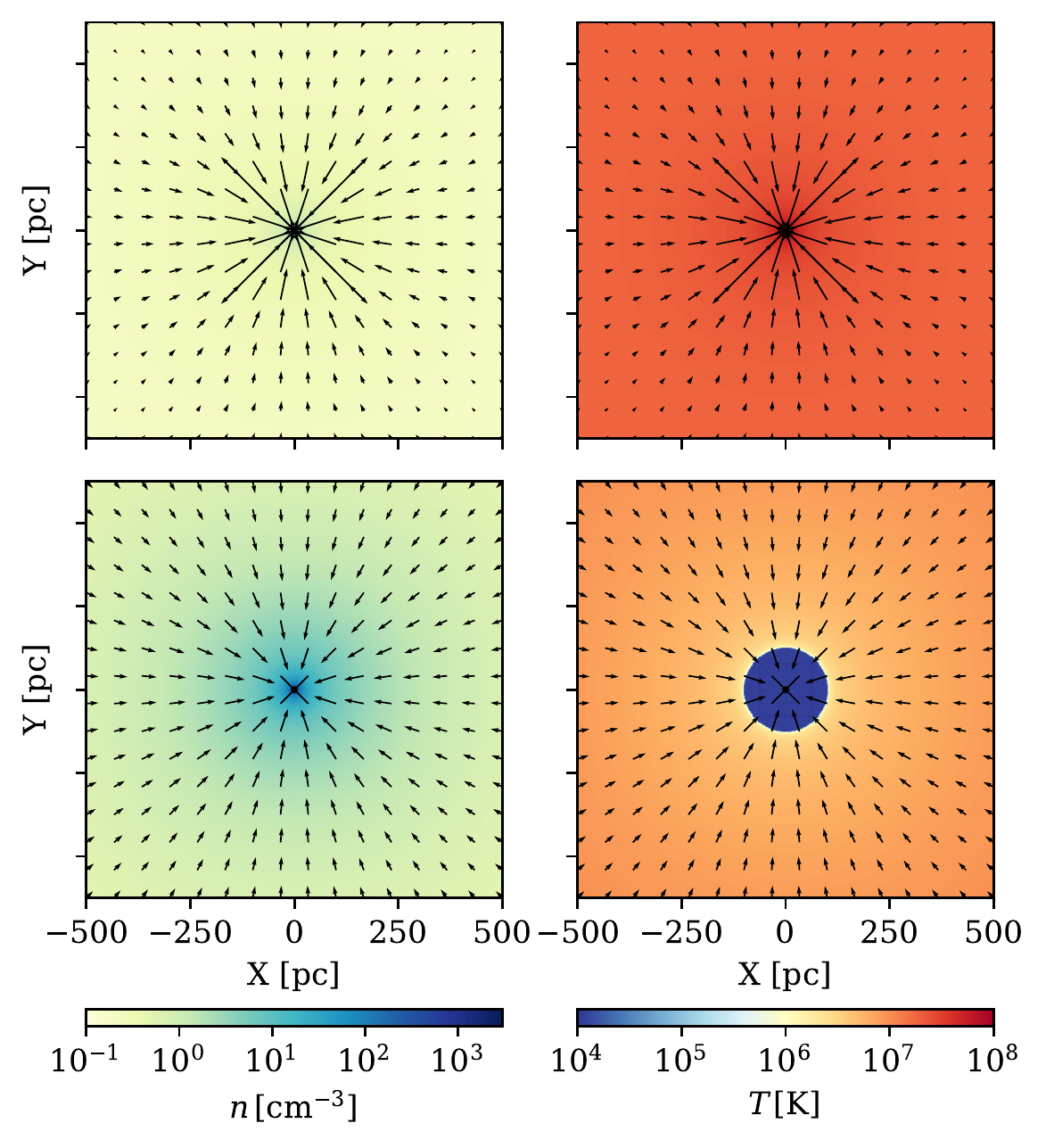}
    \caption{Slice of density (left) and temperature (right) in the plane of $z=0$ for the adiabatic (Model A13, top) and cooling (Model C13, bottom) spherically symmetric simulations. The velocity is normalized by Keplerian velocity. The are only little asymmetries due to the Cartesian mesh.\label{fig:sph_slice}}
\end{figure}

\begin{figure}[ht!]
    \centering
    \includegraphics[width=\linewidth]{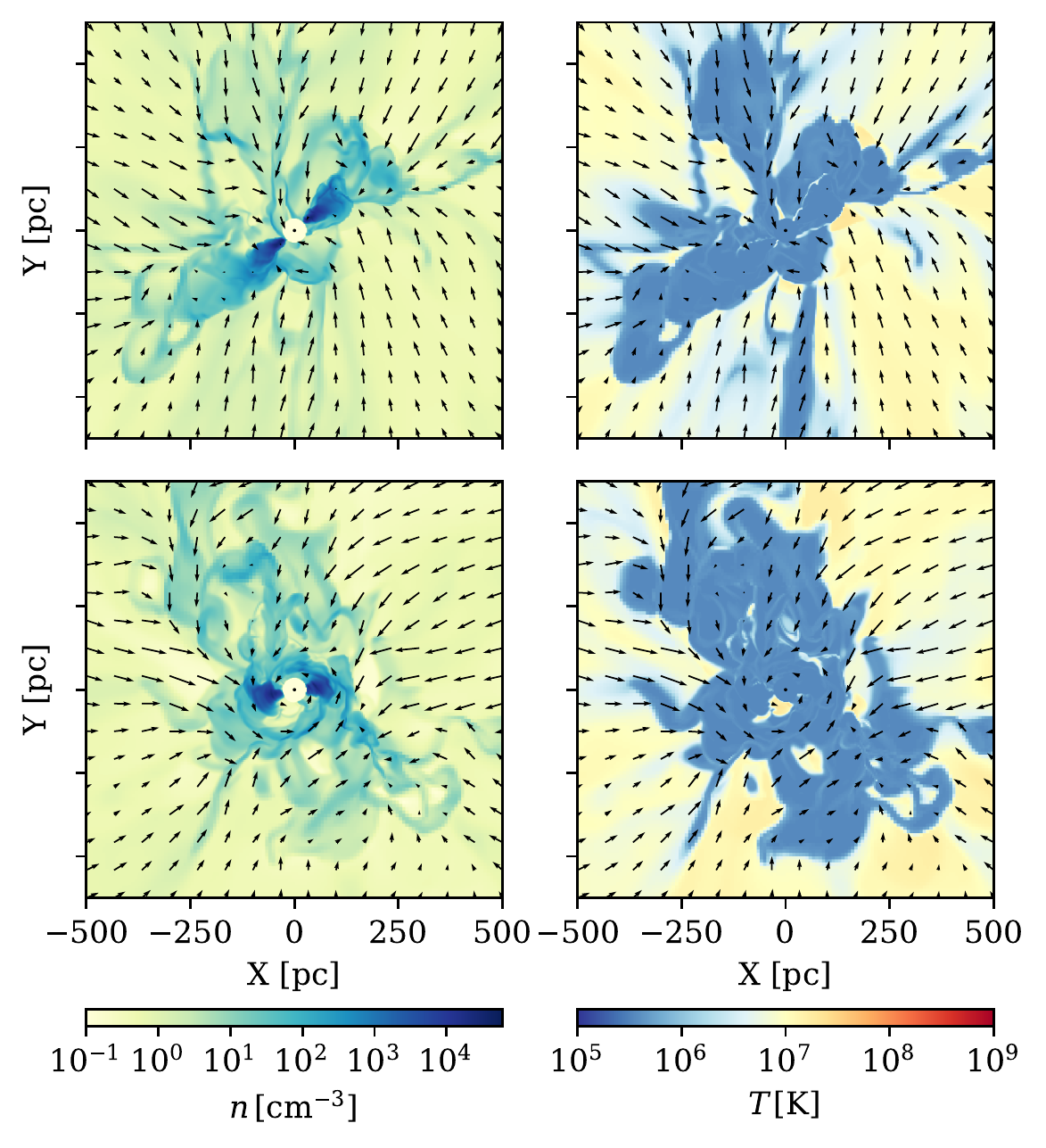}
    \caption{Similar to \figu\ref{fig:sph_slice}, but for the cooling accretion with density (top) and velocity (bottom) perturbations. The accretion flow is chaotic, but generally with a small torus in the center. There is little difference between the two types of initial perturbations. \label{fig:cool_slice}}
\end{figure}

\begin{figure}[ht!]
    \centering
    \includegraphics[width=\linewidth]{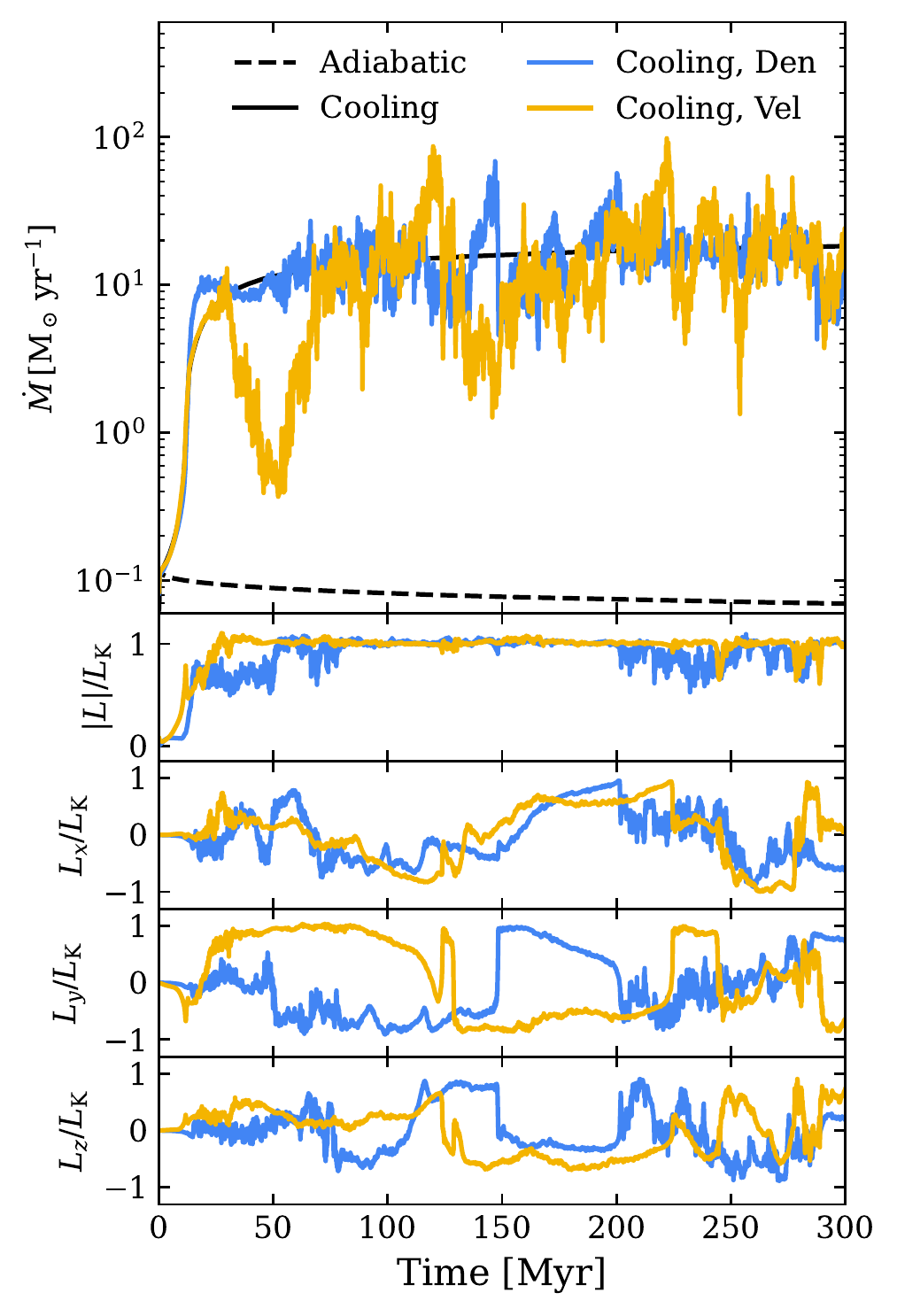}
    \caption{Time evolution of mass accretion rate (top) and normalized gas angular momentum and its three components at the inner boundary (bottom). The black lines are spherically symmetric simulations. The colored lines are cold accretion with initial density perturbation and velocity perturbation. \label{fig:cool_evo}}
\end{figure}

\subsection{Adiabatic Spherical Accretion}
First, we present our tests of adiabatic spherical accretion (Bondi accretion). For a point source, the Bondi accretion rate is
\begin{equation}
    \dot{M}_\mathrm{B}=4\pi q_\gamma\frac{G^2M^2\rho_\infty}{c_\mathrm{s,\infty}^3},
\end{equation}
where $c_\mathrm{s,\infty}$ is the sound speed at infinity, $\rho_\infty$ the density at infinity, and the factor
\begin{equation}
    q_\gamma(\gamma)=\frac{1}{4}\left(\frac{2}{5-3\gamma}\right)^{(5-3\gamma)/(2\gamma-2)},
\end{equation}
with $q_\gamma=1/4$ for $\gamma=5/3$.
At small radii, we have a self-similar solution,
\begin{eqnarray}
    u\approx-c_\mathrm{s,\infty}\left(\frac{4r}{r_\mathrm{B}}\right)^{-1/2},\ \rho\approx\rho_\infty\left(\frac{4r}{r_\mathrm{B}}\right)^{-3/2}.
\end{eqnarray}
In this study, if the gas is adiabatic, the accretion is similar to a Bondi accretion, but the potential is from the mass of a point source (the SMBH) as well as the stars and the dark matter. In addition, there is no fixed sound speed and density at infinity. The accretion rate is a function of the density and temperature of the gas, which could be different at different radii. However, due to the assumption of a flat entropy core ($\rho\propto c_\mathrm{s}^{2/(\gamma-1)}=c_\mathrm{s}^3$), the Bondi rate is constant within the core radius. 

In this test (Model A13), we employ 13 levels of mesh refinement and set $r_\mathrm{in}=0.3\,\mathrm{pc}$.
We evolve the simulation for $\sim 30$ Myr, which is about the dynamic time at 10 kpc, 100 Bondi time, and $10^6$ times the free-fall time at the inner boundary of $0.3$ pc. 
The radial profiles of density, temperature, and mass accretion rate are shown by dashed lines in \figu\ref{fig:sph_radial}. 
The accretion rate is similar to the Bondi accretion rate $\sim0.05\,M_\odot\,\mathrm{yr^{-1}}$ within $\sim 300\,\mathrm{pc}$. The radial profiles are very stable, following scaling of $\rho\propto r^{-3/2}$, $T\propto r^{-1}$ inside Bondi radius. The profiles are nearly steady over the simulation, without significant change. The density and temperature profiles are very close to the initial condition. The inflow velocities are subsonic over the simulation domain. The flat entropy core is maintained during the simulation. 
The snapshots of density and temperature are shown in \figu\ref{fig:sph_slice}. There are only tiny (nearly invisible) asymmetries due to the Cartesian mesh. 
This test builds a steady accretion flow similar to the Bondi solution in a large domain from $\sim20$ kpc down to the sub-parsec scale spanning 5 orders of magnitude.

\subsection{Cooling Spherical Accretion}

Radiative cooling drastically changes the accretion flow when adiabaticity is violated. Though the free-fall time of the gas is smaller than the gas cooling time by a factor of 5 or higher, the ratio of cooling time and gas inflow time can be smaller. As is shown in the adiabatic case, the gas cooling time is much shorter than the inflow time at $\sim$ 1 kpc. In this test (Model C13), we include radiative cooling and set a temperature floor of $T_\mathrm{floor}=10^4\,\mathrm{K}$.
The floor is different from the fiducial suite because there is no disk to resolve. Thus we can safely set a more physical temperature. Similar to the adiabatic case, we also employ 13 levels of mesh refinement and set $r_\mathrm{in}=1\,\mathrm{pc}$. 

The radial profiles are plotted by solid lines in \figu\ref{fig:sph_radial}.
With cooling, the accretion rate increases exponentially and finally saturates at $\sim 20$ Myr. It is enhanced by a factor of $\sim 100$, approaching $\sim10\,M_\odot\,\mathrm{yr^{-1}}$. The gas within $\sim 100\,\mathrm{pc}$ cools down to the temperature floor of $10^4\,\mathrm{K}$ in $\sim 20$ Myr. 
There is a cooling flow outside $\sim 300\,\mathrm{pc}$ with $t_\mathrm{cool}\sim t_\mathrm{inflow}$. 
The gas density within $\sim 1\,\mathrm{kpc}$ increases by up to two orders of magnitude, following a power law of $r^{-3/2}$ in the whole simulation domain.
The cold gas is quasi-free-fall within $\sim 100\,\mathrm{pc}$. The sonic point is at $\sim 300\,\mathrm{pc}$. 
The compressional heating does not influence the accretion flow.
For the hot accretion flow, the time ratio $t_\mathrm{cool}/t_\mathrm{ff}$ increases towards smaller radii. But if cooling develops in about one cooling time, it changes the boundary condition substantially and thus we have $t_\mathrm{cool}/t_\mathrm{ff}\lesssim 1$. The Bondi solution is not appropriate in this circumstance. 

The snapshots in \figu\ref{fig:sph_slice} show the condensation of the galactic core. The collapse is symmetric and monolithic. We do not find thermal instabilities present, though there are tiny asymmetries. The giant core continues to expand slowly in size. The flow becomes supersonic at the boundary of the cold nucleus and the hot ambient gas, consistent with the region where the gas is in a fast cooling regime ($\Lambda(T)\propto T^{-1/2}, T\lesssim10^6\,\mathrm{K}$). Even if the initial sonic point is within the Bondi radius, the mass accretion rate would be so large pushing the sonic point $r_\mathrm{s}>r_\mathrm{B}$. The solutions are in good numerical stability with very low oscillations both in space and in time. We note that there are still some small asymmetries in smaller radii.

\subsection{Turbulent Cold Accretion}

It is worthwhile to investigate the accretion flow in the existence of turbulence and cooling but no heating, which sets the upper limit of chaotic cold accretion to some extent. The tests are Models CD10 and CV10 in \tab\ref{tab:models}. The setup is identical to Models CHD10 and CHV10 except that we do not add any artificial heating. 

The snapshots and time evolution are shown in \figus\ref{fig:cool_slice} and \ref{fig:cool_evo}. The mass accretion rate is fluctuating around the spherically symmetric cooling accretion rate. We still find that there is usually a small torus within $\sim 100\,\mathrm{pc}$ with the angular momentum of the gas around the inner boundary being Keplerian. The accretion flow is more chaotic compared with the simulation with heating. The disk changes its orientation more frequently. There is little difference between the two types of initial perturbations. Due to the absence of heating, there is more cold gas formed. The accretion rate is therefore higher.

\bibliography{ref}{}
\bibliographystyle{aasjournal}

\end{CJK*}
\end{document}